\documentclass[iop]{emulateapj}

\usepackage{graphicx}
\usepackage{stmaryrd}
\usepackage{multirow}
\usepackage{natbib}
\usepackage{xspace}
\usepackage{hyperref}
\usepackage{amsmath}

\newcommand{\simi}{\ensuremath{\sim}}

\newcommand{\degree}{\ensuremath{^{\circ}}\xspace}
\newcommand{\xos}{\mbox{XO-6}\xspace}
\newcommand{\xosb}{\mbox{XO-6b}\xspace}
\newcommand{\eg}{\textit{e.g.}\xspace}

\newcommand{\kepler}{\textit{Kepler}\xspace}
\newcommand{\corot}{\textit{CoRoT}\xspace}
\newcommand{\modif}{}
\newcommand{\modifm}{}

\shorttitle{XO-6\MakeLowercase{$\rm b$}}

\shortauthors{Crouzet et al.}

\begin{document}

\title{Discovery of XO-6\MakeLowercase{b}: a hot Jupiter transiting a fast rotating F5 star \modif{on an oblique orbit}}

\author{N.~Crouzet~\altaffilmark{1}}
\author{P.~R.~McCullough~\altaffilmark{2}}
\author{D.~Long~\altaffilmark{3}}
\author{P.~Montanes~Rodriguez~\altaffilmark{4}}
\author{A.~Lecavelier~des~Etangs~\altaffilmark{5}}
\author{I.~Ribas~\altaffilmark{6}}
\author{V.~Bourrier~\altaffilmark{7}}
\author{G.~H\'ebrard~\altaffilmark{5}}
\author{F.~Vilardell~\altaffilmark{8}}
\author{M.~Deleuil~\altaffilmark{9}}
\author{E.~Herrero~\altaffilmark{6,8}}
\author{E.~Garcia-Melendo~\altaffilmark{10,11}}
\author{L.~Akhenak~\altaffilmark{5}}
\author{J.~Foote~\altaffilmark{12}}
\author{B.~Gary~\altaffilmark{13}}
\author{P.~Benni~\altaffilmark{14}}
\author{T.~Guillot~\altaffilmark{15}}
\author{M.~Conjat~\altaffilmark{15}}
\author{D.~M\'ekarnia~\altaffilmark{15}}
\author{J.~Garlitz~\altaffilmark{16}}
\author{C.~J.~Burke~\altaffilmark{17}}
\author{B.~Courcol~\altaffilmark{9}}
\author{O.~Demangeon~\altaffilmark{9}}

\affil{\altaffilmark{1} Dunlap Institute for Astronomy \& Astrophysics, University of Toronto, 50 St. George Street, Toronto, Ontario, Canada M5S 3H4}
\affil{\altaffilmark{2} Department of Physics and Astronomy, Johns Hopkins University, 3400 North Charles Street, Baltimore, MD 21218, USA}
\affil{\altaffilmark{3} Space Telescope Science Institute, 3700 San Martin Dr, Baltimore, MD 21218, USA}
\affil{\altaffilmark{4} Instituto de Astrof\'isica de Canarias, C/V\'ia L\'actea s/n, E-38200 La Laguna, Spain}
\affil{\altaffilmark{5} Institut d'Astrophysique de Paris, UMR7095 CNRS, Universit\'e Pierre \& Marie Curie, 98bis boulevard Arago, 75014 Paris, France}
\affil{\altaffilmark{6} Institut de Ci\`encies de l'Espai (CSIC-IEEC), Campus UAB, Carrer de Can Magrans s/n, 08193 Bellaterra, Spain}
\affil{\altaffilmark{7} Observatoire de l'Universit\'e de Gen\`eve, 51 chemin des Maillettes, 1290 Sauverny, Switzerland}
\affil{\altaffilmark{8} Observatori del Montsec (OAdM), Institut d'Estudis Espacials de Catalunya (IEEC), Gran Capit\`a, 2-4, Edif. Nexus, 08034 Barcelona, Spain}
\affil{\altaffilmark{9} Aix Marseille Universit\'e, CNRS, LAM (Laboratoire d'Astrophysique de Marseille) UMR 7326, 13388 Marseille, France}
\affil{\altaffilmark{10} Departamento de F\'isica Aplicada I, Escuela T\'ecnica Superior de Ingenier\'ia, Universidad del Pa\'is Vasco UPV/EHU, Alameda Urquijo s/n, 48013 Bilbao, Spain}
\affil{\altaffilmark{11} Fundacio Observatori Esteve Duran, Avda. Montseny 46, Seva E-08553, Spain}
\affil{\altaffilmark{12} Vermillion Cliffs Observatory, 4175 E. Red Cliffs Drive, Kanab, UT 84741, USA}
\affil{\altaffilmark{13} Hereford Arizona Observatory, 5320 East Calle de la Manzana, Hereford, AZ 85615, USA}
\affil{\altaffilmark{14} Acton Sky Portal, Acton, MA, USA}
\affil{\altaffilmark{15} Laboratoire Lagrange, Universit\'{e} C\^ote d'Azur, Observatoire de la C\^ote d'Azur, CNRS, Boulevard de l'Observatoire, CS 34229, F-06304 Nice Cedex 4, France}
\affil{\altaffilmark{16} Elgin Observatory, Elgin, OR, USA}
\affil{\altaffilmark{17} SETI Institute/NASA Ames Research Center, Moffett Field, CA 94035, USA}

\email{crouzet@dunlap.utoronto.ca}

\begin{abstract}

Only a few hot Jupiters are known to orbit around fast rotating stars. These exoplanets are harder to detect and characterize and may be less common than around slow rotators. Here, we report the discovery of the transiting hot Jupiter \xosb, which orbits a bright, hot, and fast rotating star: V = 10.25, $T_{eff\star} = 6720 \pm 100$ K, $v\,$sin$\,i_{\star} = 48 \pm 3 \, \rm km\,s^{-1}$. We detected the planet from its transits using the XO instruments and conducted a follow-up campaign. Because of the fast stellar rotation, radial velocities taken along the orbit do not yield the planet's mass with a high confidence level, but we secure a 3-$\sigma$ upper limit $M_p < 4.4 \; \rm M_{Jup}$. We also obtain high resolution spectroscopic observations of the transit with the SOPHIE spectrograph at the 193-cm telescope of the Observatoire de Haute-Provence and analyze the stellar lines profile by Doppler tomography. The transit is clearly detected in the spectra. The radii measured independently from the tomographic analysis and from the photometric lightcurves are consistent, showing that the object detected by both methods is the same and indeed transits in front of \xos. We find that XO-6b lies on a prograde and misaligned orbit with a sky-projected obliquity $\modifm{\lambda=-20.7\degr \pm 2.3 \degr}$. The rotation period of the star is shorter than the orbital period of the planet: $P_{rot} < 2.12$ days, $P_{orb} = 3.77$ days. Thus, this system stands in a largely unexplored regime of dynamical interactions between close-in giant planets and their host stars.

\end{abstract}


\section{Introduction}

Most of our understanding of close-in gas giant planets is enabled by studying those transiting in front of bright stars. Such systems are favorable to photometric and radial velocity observations and to atmospheric characterization. The vast majority of these systems have been discovered by ground-based surveys such as WASP \citep{Pollacco2006, Collier2007} and HAT \citep{Bakos2004}. These two projects detected a total of \simi60 hot Jupiters around stars of magnitude $J < 10$. The \corot and \kepler missions provided little addition to this sample because they target fainter stars ; \kepler discovered two planets around stars of magnitude $J < 10$ with a radius larger than 0.3 R$\rm_{Jup}$ : Kepler-16b and Kepler-25c. \modif{The ratio of such hot Jupiters numbers (60:2) is approximately the same as the corresponding ratio of survey areas on the sky (30,000:100 square degrees).}

Among parameters used to characterize hot Jupiters, the distribution of obliquities draws particular attention. Sky-projected obliquities can be measured by observing exoplanets in spectroscopy during a transit: the planet occults a portion of its rotating host star and distorts the apparent stellar line profile ; this effect is known as the Rossiter-McLaughlin effect \citep{Holt1893, Rossiter1924, McLaughlin1924}. It is most commonly analyzed through velocimetry. A more sophisticated technique, Doppler tomography, has also been used in a few cases and provides more precise parameters as well as a wider range of information on the planet and the host star \citep[\eg][]{Collier2010a, Collier2010b, Bourrier2015}. 
\nolinebreak
With the rapid growth of Rossiter-McLaughlin measurements in recent years, obliquities have become a powerful way to probe different theories for the dynamical history of hot Jupiters. \citet{Winn2010a} noted that hot Jupiters orbiting stars with relatively cool photospheres ($\modifm{T < 6250}$ K) have low obliquities whereas those orbiting hotter stars show a wide range of obliquities. This trend has been investigated in the context of dynamical interactions \citep[\eg][]{Schlaufman2010, Albrecht2012, Dawson2014, Winn2015}.
\nolinebreak
The comparison of planets' orbital periods and host stars' rotation periods also constrains the dynamics of hot Jupiters \citep{Mazeh2005,Pont2009,Husnoo2012}. However, the vast majority of known hot Jupiters orbit around slowly rotating stars. This trend is supported by a dearth of \kepler Objects of Interest at short orbital periods around fast rotating stars \citep{McQuillan2013} ; such studies could be used to calibrate potential selection effects of ground-based transit surveys against detecting planets around fast rotating stars.

The XO project \citep{McCullough2005} aims at detecting transiting exoplanets around bright stars from the ground with small telescopes. The project started in 2005 and discovered five close-in gas giant planets, XO-1b to XO-5b\footnote{\citet{McCullough2006, Burke2007, JohnsKrull2008, McCullough2008, Burke2008}}. A new version of XO was deployed in 2011 and 2012 and operated nominally from 2012 to 2014. In this paper, we report the discovery of \xosb, a transiting hot Jupiter orbiting a bright and fast rotating star. After its detection by the XO instruments, we conducted a follow-up campaign via photometry, radial velocity, and Rossiter-McLaughlin observations including a Doppler tomography analysis in order to establish the presence of the planet and extract the system parameters.   

The XO setup and observations are described in Section \ref{sec: Observations} and the data reduction pipeline in Section \ref{sec: Data reduction}. The follow-up campaign and the nature of the \xos system are presented in Section \ref{sec: Follow-up campaign}. The stellar parameters are derived in Section \ref{sec: Stellar parameters}. In Section \ref{sec: Discussion}, we discuss the properties of the \xos system in the context of dynamical studies of hot Jupiters and their host stars, and we conclude in Section \ref{sec: Conclusion}.

\section{Discovery Instrumentation and Observations}
\label{sec: Observations}

The new XO instrumental setup consists of three identical units installed at Vermillion Cliffs Observatory, Kanab, Utah, at Observatorio del Teide, Tenerife, Canary Islands, and at Observatori Astron\`omic del Montsec (OAdM), near \`Ager, Spain. Each unit is composed of two 10 cm diameter and 200 mm focal length Canon telephoto lenses equipped with an Apogee E6 $1024 \times 1024$ pixels CCD camera mounted on a German-Equatorial Paramount ME mount and protected by a shelter with a computer-controlled roof. All six lenses and cameras operate in a network configuration and point towards the same fields of view. These fields of view are different from those of the original XO, with no overlap. Instrumental parameters and data reduction methods are similar as in \citet{McCullough2005} unless indicated. The focus is adjusted to yield a PSF (Point Spread Function) FWHM (Full Width Half Maximum) of \simi 1 pixel; in practice the FWHM varies between 1 and 1.5 pixels. The CCDs are used in spatial-scan mode: pixels are read continuously while stars move along on the detector. The mount tracking and CCD reading rates are calculated to yield round PSFs. The resulting images are long strips of $43.2^{\circ} \times 7.2^{\circ}$ instead of $7.2^{\circ} \times 7.2^{\circ}$ images that would result if the CCD was used in staring mode. This technique reduces overheads, increases the time spent at collecting photons, and maximizes the number of observed bright stars. The same two strips of the sky were observed over and over. Between 50 and 100 strips were recorded each night by each camera; the exposure time is 317 s for a full strip (53 s for each $7.2^{\circ} \times 7.2^{\circ}$ region). A weather sensor records meteorological data such as the ambient temperature, sky temperature, humidity, rain drops, wind speed and direction, dew point, etc... every minute. Each unit operates robotically: the weather data are interpreted in real time, a command is sent to open or close the roof, and run the observations. The units can also be controlled remotely. A rotating webcam sensitive in the visible and near-infrared is mounted inside each shelter and gives a live view of the systems. The observations span a duration of two times nine months between 2012 and 2014, with gaps due only to weather and instrumental problems.

\section{Data reduction}
\label{sec: Data reduction}

First, data taken by the six camera systems are reduced independently. The strips are carved into $1024 \times 1024$ pixel images, and we discard the first $1024 \times 1024$ image of each strip where the scanning is just starting. This yields 9 fields of $7.2^{\circ} \times 7.2^{\circ}$ ; the field located around the celestial North pole is covered by both strips and has the maximum phase coverage. An approximate WCS (World Coordinate System) solution is found for each image using the astrometry.net software program\footnote{\url{http://astrometry.net/}} \citep{Lang2010}, and it is improved using a 6 parameter astrometric solution. 

Dark frames are taken at the beginning and end of each night and are averaged to yield one dark per night. A flat-field is built from twilight flats and is used for the whole duration of the observations. Science frames are calibrated by the dark and flat. Warm columns are identified using high-pass filters; columns with a excess of 15 ADU with respect to their 10 neighbouring columns are flagged and their excess flux is removed. We find typically between 0 and 5 warm columns per image, except for one CCD which has up to 100 warm columns.

We select the target stars from a reference image taken under very good conditions. This yields about 6000 stars in each field. To identify the target stars on each image, the coordinates of point sources are correlated to that of the target stars in the reference image. We perform circular aperture photometry using the Stellar Photometry Software program \citep{Janes1993}. The sky background is calculated in an annulus around each star and is subtracted. The optimum photometric aperture radius is computed as a function of stellar magnitude for each camera using data taken under average conditions, and is kept constant for each star; this radius varies from 2 to 10 pixels depending on the star magnitude. The photometric measurements are gathered in a star-epoch array where stars are sorted by ascending instrumental magnitude. One array is obtained for each field and each camera. 

The next reduction steps are performed only on the 2000 brightest stars of each field, which corresponds approximately to a limit magnitude V = 12 and a 1\% RMS photometric precision. Several calibrations are applied independently for each scan direction (the strips can be observed scanning North or South). The mean magnitude of each star is subtracted; we use the magnitude residuals in all the following. \modif{We select the reference stars for each star and each scan direction separately. For a given star $s$, we subtract the photometric time series of $s$ from those of all the stars, and evaluate the mean absolute deviation (MAD) of the resulting time series. Stars are sorted by increasing MAD; the first index is excluded because it corresponds to star $s$, and the following N stars are kept as reference stars. Then, we build a reference time-series using an outlier resistant mean of the N reference stars, and subtract it to the time-series of star $s$. This yields a calibrated lightcurve for star $s$. We use N = 10 for all the stars.} Then, for each epoch, we remove a 3rd oder polynomial corresponding to low-frequency variations of magnitude residuals in the CCD's \textit{x,y} plane, also called ``L-flats". We also remove a linear dependance of each lightcurve with airmass.

Bad data are flagged according to several criteria: the quality of each epoch measured from the dispersion of the residuals of a subset of stars, the sky brightness and its homogeneity (which account for clouds or a shelter's wall partially obscuring the field of view), the airmass, the number of detected stars, and the image cross-correlation (revealing for example double point sources due to an imperfect tracking occurring occasionally on one of the mounts). Then, systematic effects are removed using the Sysrem algorithm \citep{Tamuz2005}. Finally, the lightcurves from the 6 cameras are combined together and we search for periodic signals using the BLS (Box Least Square) algorithm \citep{Kovacs2002}. We keep signals compatible with planetary transits for visual inspection and detailed investigation. 

\begin{figure}[htbp]
   \centering
   \includegraphics[width=8cm]{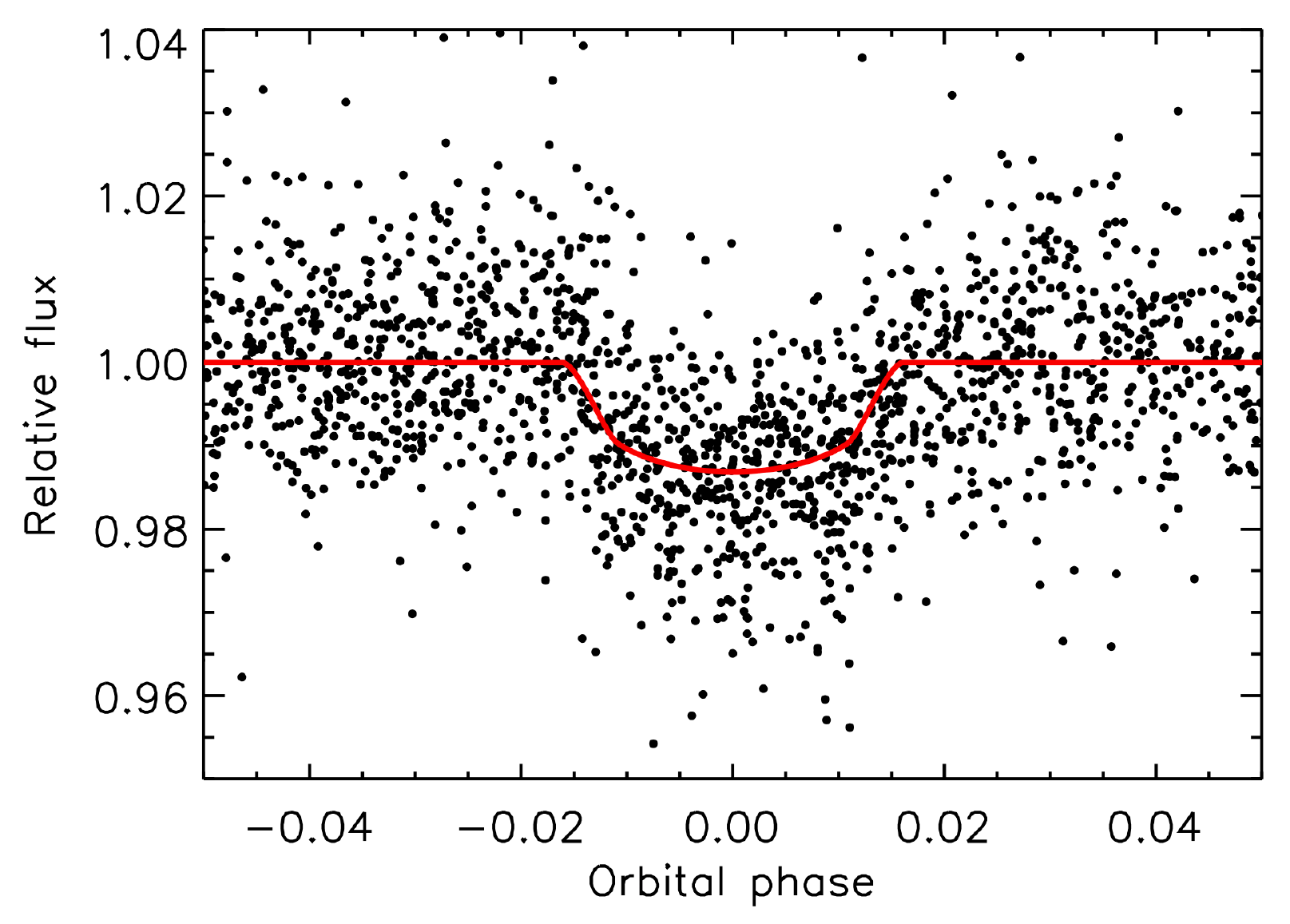}
      \caption{Discovery lightcurve of \xosb (black dots) and best transit fit using the parameters from Table~\ref{tab: final parameters} (red line).}
   \label{fig: xo lightcurve}
\end{figure}

The discovery lightcurve of \xosb is shown in Figure~\ref{fig: xo lightcurve}. These data were obtained from the three units (six cameras) of the new XO instruments during the first campaign, from fall 2012 to spring 2013.

\section{Follow-up campaign}
\label{sec: Follow-up campaign}

We conducted follow-up observations and analyses in photometry and spectroscopy to establish the nature of the system. The first spectroscopic measurements reveal that the star is a fast rotator, which jeopardizes measuring the mass of the companion by radial velocities (see Section \ref{sec: Radial velocity follow-up}). Therefore, particular care is given in the photometric follow-up analysis to look for signs of a stellar eclipsing binary or a triple star system, as detailed in the following. 
Although the XO lightcurve is contaminated by another star which is 1 magnitude fainter and located at 38'' separation from XO-6, both stars are well resolved by the follow-up instruments. \modif{Nine other stars are located within 1 arcminute of XO-6 and are 4.5 to 7 magnitude fainter; they do not affect the follow-up measurements.}

\subsection{Photometric follow-up}
\label{Photometric follow-up}

The photometric follow-up observations were conducted by an extended team of amateur and professional astronomers with facilities summarized in Table \ref{tab: photometric follow-up}. We observed 18 transit events with good quality data, from December 26th, 2013 to February 2nd, 2015 using different filters (g', r', i', B, V, R, I); some transits were observed alternating between two filters. In total, this yields 26 sets of individual transits and filters that we use in the analysis (Figure \ref{fig: lc individual transits}).

\begin{table*}
\begin{center}
\caption{Observatories and telescopes used for the photometric follow-up. \\ The diameters of the primary mirrors are given in inches and cm.}
\label{tab: photometric follow-up}
\begin{tabular}{lll}
\hline
\hline
Observatory & Telescope & Label \\
\tableline 
Hereford Arizona Observatory, Arizona, USA & Celestron 11" (28 cm) - Meade 14" (36 cm) & HAO \\
Acton Sky Portal, Massachusetts, USA & 11" (28 cm)  & ASP \\
Observatori Astron\`omic del Montsec, Catalonia, Spain & Joan Or\'o Telescope 31" (80 cm)  &  TJO \\
Observatoire de Nice, France & Schaumasse 16" (40 cm)  & NCE \\
Vermillion Cliffs Observatory, Kanab, Utah, USA & 24'' (60 cm)  &  VCO \\
Elgin Observatory, Elgin, Oregon, USA & 12" (30 cm)  &  EO \\
\hline
\hline
\end{tabular}
\end{center}
\end{table*}

\subsubsection{Analysis of individual transits}
\label{Analysis of individual transits}

\begin{figure}[htbp]
   \centering
   \includegraphics[width=7.8cm]{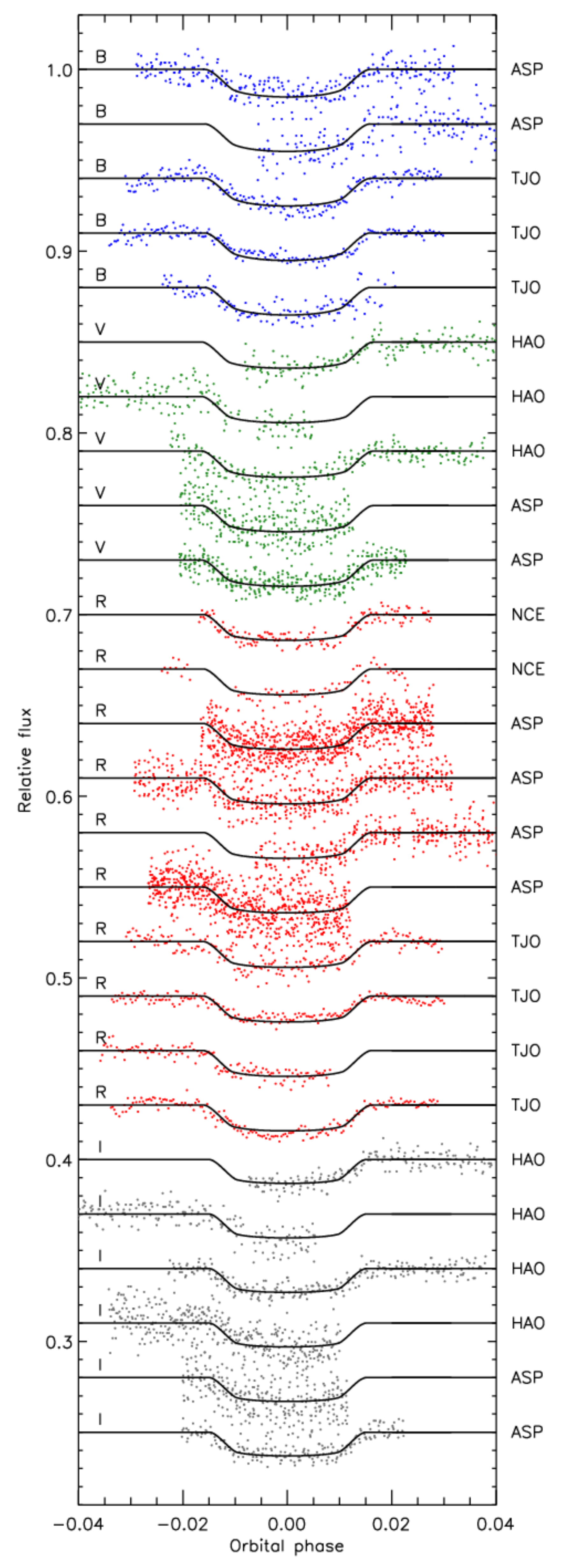}
      \caption{Photometric follow-up observations of XO-6b for individual transits and bandpasses. The bandpasses are noted as B (blue), V (green), R (red), and I (grey), and the observatories are labeled as in Table \ref{tab: photometric follow-up}. Only lightcurves of good quality are shown. The best transit model calculated for each bandpass is overplotted as a black line. \modif{The different lightcurves do not have the same time sampling, therefore the apparent point-to-point dispersion is not representative of their relative quality}. Lightcurves are offset for clarity.}
   \label{fig: lc individual transits}
\end{figure}

First, each transit time series is analyzed individually in order to search for signs of false-positive configurations (such as differences between even and odd transits or large transit timing variations) and to refine the ephemeris. Julian dates are converted to BJD (Barycentric Julian Dates). Each lightcurve is fitted by a combination of a transit model from \citet{Mandel2002} and a linear trend to correct for systematic effects. We assume a circular orbit. We use the downhill simplex minimization procedure with the following parameters: $P$, $T_0$, $a/R_{\star}$, $R_p/R_{\star}$, $i$, and $t_0$, where $P$ is the orbital period, $T_0$ the average mid-transit time, $a$ the semi-major axis, $R_{\star}$ the stellar radius, $R_p$ the planetary radius, $i$ the inclination, and $t_0$ the offset of the mid-transit time of each transit with respect to $T_0$. We include two more parameters to account for a flux offset and a linear trend. We calculate the limb-darkening coefficients using the John Southworth's JKTLD software program \citep{Southworth2015}\footnote{\url{http://www.astro.keele.ac.uk/jkt/codes/jktld.html}} with the \citet{Claret2000, Claret2004} tables, the ATLAS model for the stellar atmosphere, a quadratic limb-darkening law, and the stellar parameters $T_{eff\star} = 6720$ K, log$g_\star$ = 4.036, $M/H = 0$, and $V_{micro} = 2 \; \rm km\,s^{-1}$ (see Section \ref{sec: Stellar parameters}). We fix these coefficients to their theoretical values in each bandpass. We remove outliers lying away from the transit model by more than 3 $\sigma$, where $\sigma$ is the standard deviation of the lightcurve. The individual lightcurves are displayed in Figure \ref{fig: lc individual transits}.

We improve the transit ephemeris parameters $P$ and $T_0$ in an iterative process. First, $t_0$ is fixed to 0 whereas $a/R_{\star}$, $R_p/R_{\star}$, and $i$ are free for each transit. Then, $a/R_{\star}$, $R_p/R_{\star}$, and $i$ are fixed to their mean over all transits and $t_0$ is allowed to vary. Finally, we calculate a linear fit between the set of individual $t_0$ and the period index, and we update $P$ and $T_0$ in order to remove this linear variation. This new ephemeris as well as the new values of $a/R_{\star}$, $R_p/R_{\star}$, and $i$ are used as an initial guess for the next iteration. We perform 100 iterations. The final distributions for the parameters are well behaved, and we use the median values of the distributions of $P$ and $T_0$ as the final ephemeris. We compute the uncertainties in the ephemeris in two ways. A first estimate is obtained from the standard deviation of these $P$ and $T_0$ distributions. A second estimate is obtained using the \modif{residual-permutation method \citep[see][and references therein]{Pont2006}}, in which we shift the $t_0$ residuals with respect to the period index, calculate a new linear fit, and derive a new ephemeris. Applying all possible shifts results in distributions for $P$ and $T_0$, and their standard deviations yield the uncertainties on these parameters, which are larger than the first estimate by a factor of 1.2 and 1.5 for $P$ and $T_0$ respectively. We keep the larger values, obtained from the \modif{residual-permutation} method, as the final uncertainties. The final values and uncertainties of $P$ and $T_0$ are reported in Table \ref{tab: final parameters}. This ephemeris is in excellent agreement with the original XO lightcurve. In this analysis, we find no evidence for transit timing variations and no significant difference between odd and even transits at the level of precision of our data. 

\subsubsection{Analysis of transits per bandpass}
\label{Analysis of transits per bandpass}

\begin{figure}[htbp]
   \centering
   \includegraphics[width=8cm]{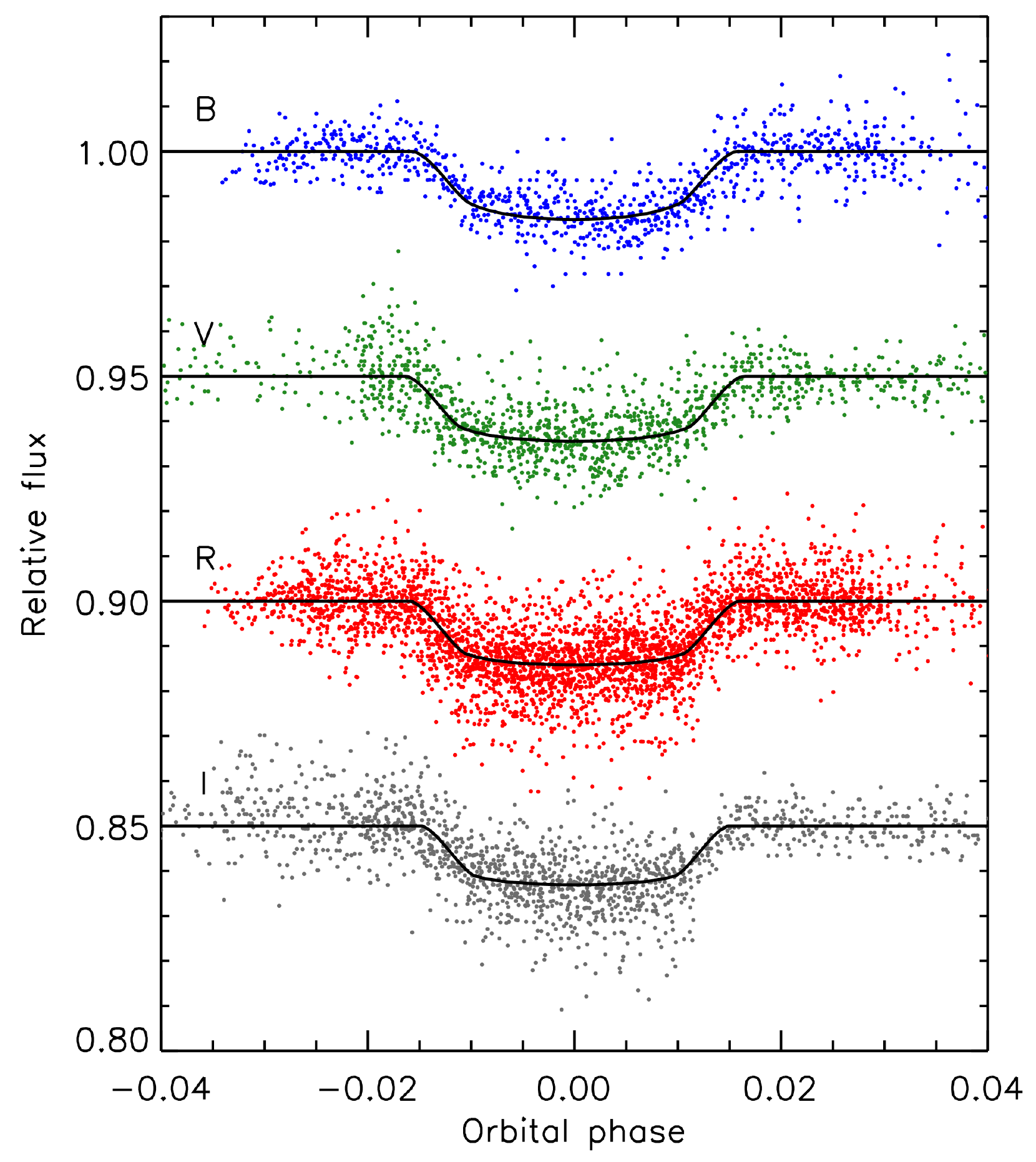}
      \caption{Photometric follow-up observations of XO-6b in several bandpasses: B (blue), V (green), R (red), and I (grey) from top to bottom. \modif{Data from different observations are blended in their respective filter bands}. The best transit model in each bandpass is overplotted as a black line. Lightcurves are offset for clarity.}
   \label{fig: lc filters}
\end{figure}

We gather data and fit a transit model in each bandpass (B, V, R, I). We fix $P$ and $T_0$ to the values derived in Section \ref{Analysis of individual transits}, and the limb-darkening coefficients to their theoretical values in each bandpass. The free parameters are $a/R_{\star}$, $R_p/R_{\star}$, and $i$, plus a flux offset and a linear trend. We assume a circular orbit and use the downhill simplex minimization procedure. Then, the uncertainties are calculated using a customized \modif{residual-permutation} method: instead of shifting the residuals over the whole transit curve, we divide this curve into two parts from mid-transit, and for each shift on one side all shifts are successively applied on the other side. This results in $\mathcal{O}(N^2)$ permutations instead of $\mathcal{O}(N)$, where $N$ is the number of data points, and allows correlated noise features to be shuffled with respect to each other before and after the transit. The best transit model is calculated for each permutation using a Levenberg-Marquardt minimization method, and the standard deviations of the posterior distributions yield the 1-$\sigma$ uncertainties on the transit parameters. These parameters are consistent between the different bandpasses within their uncertainties (Table \ref{tab: fit bandpass}, Figure \ref{fig: lc filters}). We obtain the final transit parameters by a weighted average of all bandpasses with weights of $1/ \sigma^2$ where $\sigma$ is the uncertainty in each bandpass.

\begin{table}[htdp]
\begin{center}
\caption{Transit parameters of the XO-6 system obtained in each photometric bandpass and their weighted average.}
\label{tab: fit bandpass}
\begin{tabular}{cccc}
\hline
\hline
   Bandpass &  $a/R_{\star}$  & $R_p/R_{\star}$ & $i$ [deg]   \\
\hline

B  & 9.21 $\pm$ 0.48  &  0.1179 $\pm$ 0.0026  &  85.94 $\pm$ 0.51   \\
V  &  9.01 $\pm$ 0.35  &  0.1151 $\pm$ 0.0017  &  86.02 $\pm$ 0.39   \\
R  &  9.07 $\pm$ 0.41  &  0.1153 $\pm$ 0.0019  &  85.90 $\pm$ 0.42   \\
I   &  9.49  $\pm$ 0.36  &  0.1114 $\pm$ 0.0017  &  85.94 $\pm$ 0.33   \\
Average &  9.20 $\pm$  0.19  &  0.1144 $\pm$ 0.0013  &   85.95 $\pm$ 0.20 \\
\hline  
\hline  
\end{tabular}
\end{center}
\end{table}

\subsection{Secondary eclipse}
\label{sec: Secondary eclipse}

We do not detect the secondary eclipse in the XO lightcurve, but we can constrain the eclipse depth $\delta_e$ in the XO bandpass assuming a circular orbit. First, we calculate an upper limit on $\delta_e$ as $\sigma/\sqrt{N}$, where $\sigma$ and $N$ are the standard deviation and number of in-eclipse points respectively: we find $\delta_e < 0.114\%$ at 3-$\sigma$. However, this calculation assumes that only white noise is affecting the XO lightcurve. In a second approach, we estimate the noise from the data themselves. We eliminate the in-transit points, fold the lightcurve at $10^4$ different periods ranging from 3 to 4.5 days, split each lightcurve into segments corresponding to the eclipse duration, and calculate the mean flux in each segment. This yields a distribution that represents the flux variations over durations equivalent to that of the eclipse. The 3-$\sigma$ values of this distribution yield the 3-$\sigma$ upper limit on $\delta_e$. By this method, we find an upper limit $\delta_e < 0.175\%$ at 3-$\sigma$. This value is larger than in the case of pure white noise because it accounts for correlated noise. We also observed this system at the predicted time of an eclipse with the 36 cm telescope. We did not detect the eclipse, but the limit on $\delta_e$ obtained from these observations is less constraining than that obtained from the phase-folded XO lightcurve. Overall, we do not find any sign that would indicate a stellar eclipsing binary or triple star system.

\subsection{Radial velocity follow-up}
\label{sec: Radial velocity follow-up}

\begin{figure}[htbp]
   \centering
   \includegraphics[width=8cm]{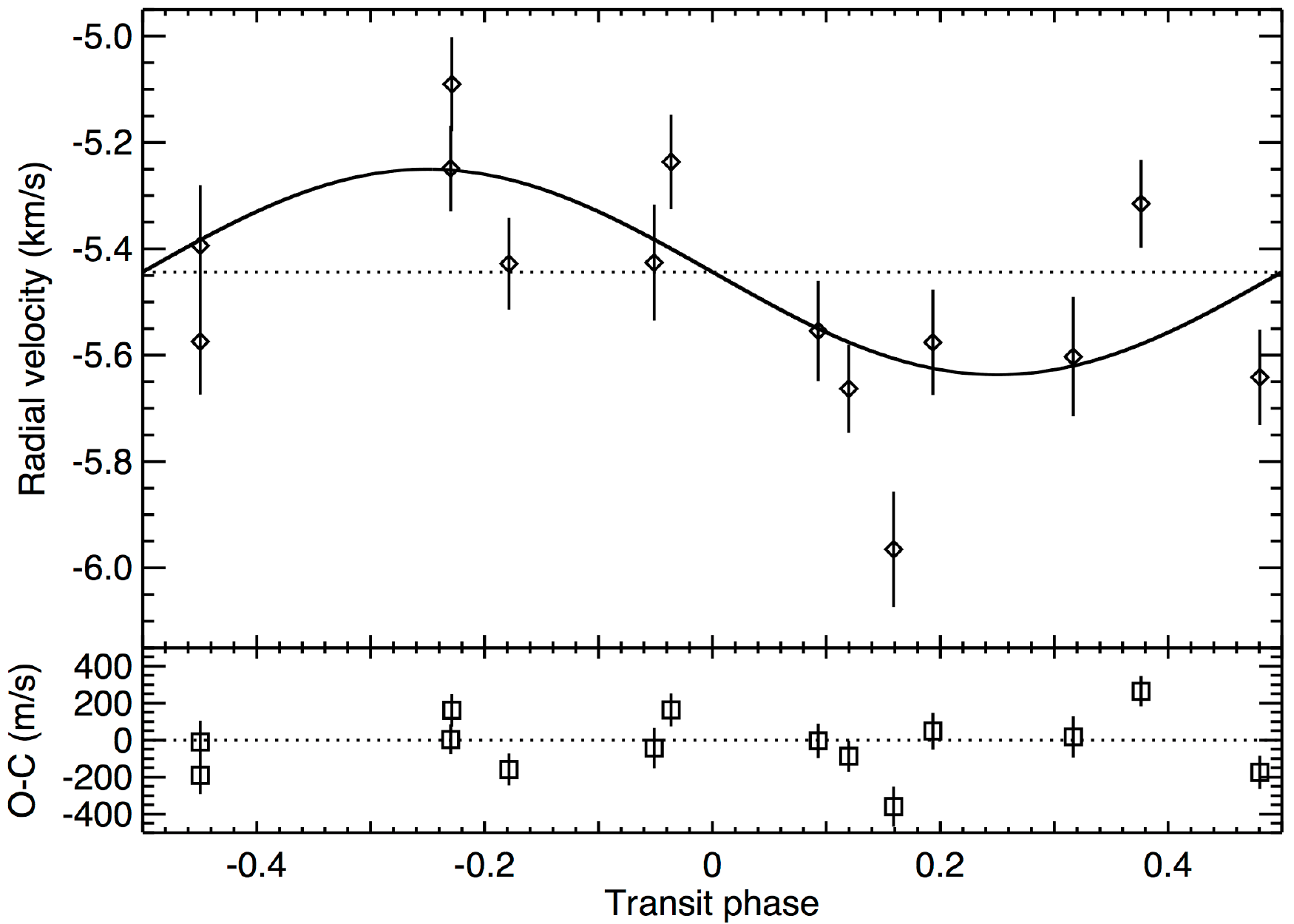}
      \caption{SOPHIE radial velocities of \xos with 1-$\sigma$ error bars, phase-folded with the period $P = 3.765$ days. A Keplerian, circular fit is overplotted as a plain line. There is a hint of detection, with a semi-amplitude $K = 200 \pm 70 \: \rm m\,s^{-1}$ corresponding to a planet mass $M_p = 1.9 \pm 0.5 \;\rm M_{Jup}$ ($K < 450 \: \rm m\,s^{-1}$ at 3 $\sigma$). The bottom panel shows the residuals of that fit.}
   \label{fig: xo6 RVs}
\end{figure}

Radial velocity follow-up was conducted \modif{between September 2013 and January 2016} with the SOPHIE spectrograph \citep{Bouchy2009} at the Observatoire de Haute-Provence, France (Table \ref{tab: xo6 RVs}). Exposure times of a few minutes allowed signal-to-noise ratios per pixel at 550 nm around 45 to be reached on most of the exposures. We used the SOPHIE pipeline to extract the spectra from the detector images, cross-correlate them with a G2-type numerical mask which produced clear cross-correlation functions (CCFs), then fit the CCFs by Gaussians to get the radial velocities \citep{Baranne1996, Pepe2002}. Measurements obtained along the orbit show that \xos is a single, fast rotating star with $v\,$sin$\,i_{\star} \approx45 \: \rm km\,s^{-1}$. The large width of the spectral lines implies a moderate accuracy of the radial velocity measurements, of the order of $\pm \:90 \: \rm m\,s^{-1}$. The phase-folded radial velocities secured out of the transit are plotted in Figure \ref{fig: xo6 RVs}. They show a hint for a detection of the reflex motion of the star due to its planet. The fit of these radial velocities with a Keplerian, circular orbit phased according to the transits provides a semi-amplitude $K = 200 \pm 70 \; \rm m\,s^{-1}$, corresponding to a planetary mass $M_p = 1.9\pm \: 0.5 \; \rm M_{Jup}$. The dispersion of the residuals is $150 \; \rm m\,s^{-1}$, slightly above the expected error bars on radial velocities, which indicates that systematic effects remain. We rather adopt the 3-$\sigma$ upper limit for its detection, $K < 450 \; \rm m\,s^{-1}$, corresponding to a planetary mass $M_p < 4.4 \; \rm M_{Jup}$. Many additional radial velocity measurements would be required for a significant measurement of the planetary mass.

\begin{figure}[htbp]
   \centering
   \includegraphics[width=8cm]{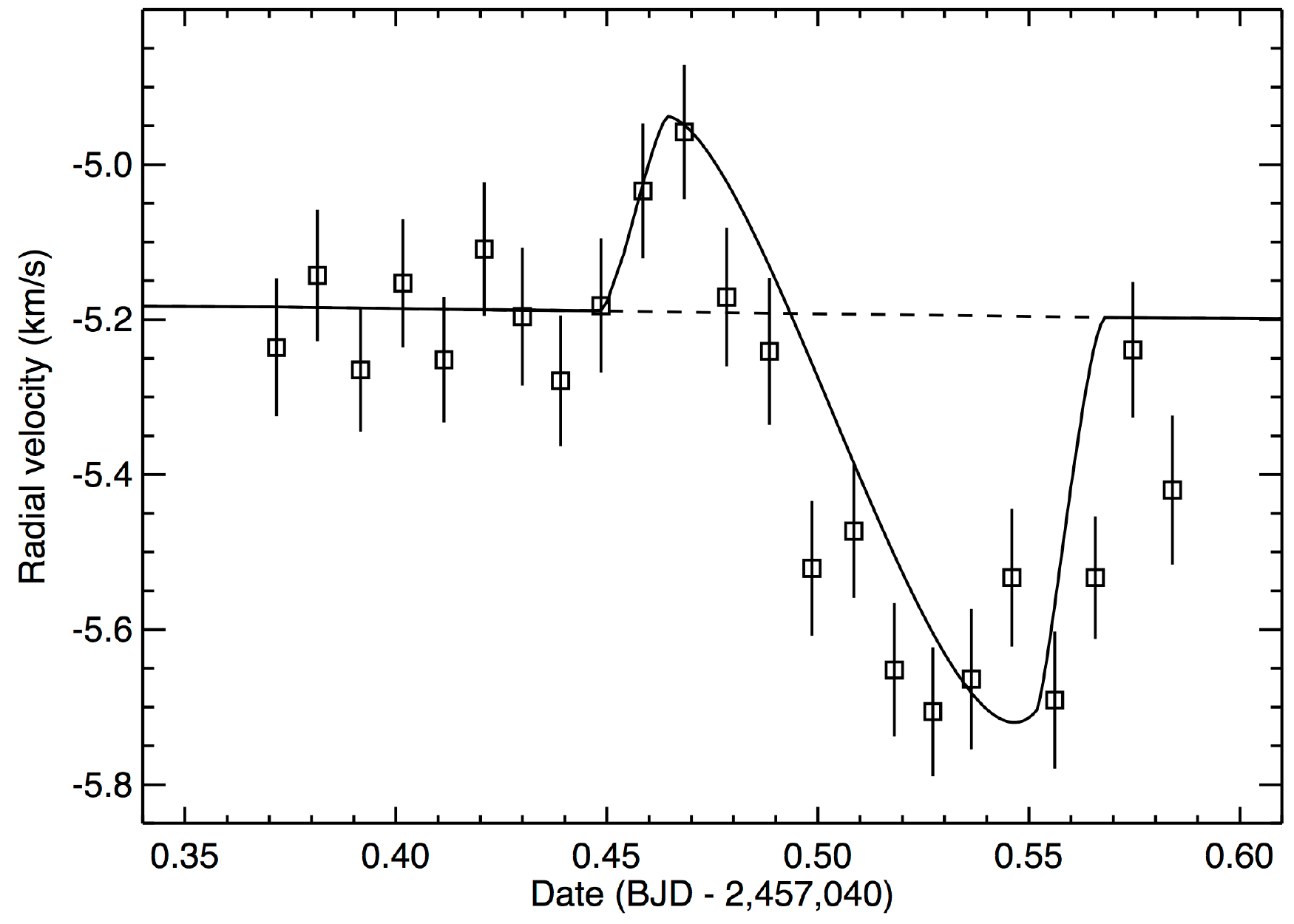}
      \caption{SOPHIE radial velocities of \xos with 1-$\sigma$ error bars, secured during the January 17, 2015 transit. The Rossiter-McLaughlin anomaly is clearly detected. Overplotted is the model corresponding to the parameters measured through Doppler tomography analysis (Section \ref{sec: Doppler tomography analysis}).}
   \label{fig: xo6 rossiter}
\end{figure}

We observed \xos with SOPHIE during a transit in order to detect and characterize the planet through the analysis of the Rossiter-McLaughlin effect. The first 6 attempts were canceled due to bad weather. Such observations require excellent weather conditions for much of a night, and a 7th attempt was made on January 17, 2015 and was successful: the transit was observed in good conditions yielding 23 measurements with signal-to-noise ratios similar as above. We clearly detect the Rossiter-McLaughlin anomaly which confirms the existence of a transiting object in front of the fast rotating star (Figure~\ref{fig: xo6 rossiter}). The amplitude is $\sim500 \: \rm m\,s^{-1}$ and the variations suggest a prograde, misaligned planetary orbit. \modif{We note that the systemic radial velocity is different by about $200 \; \rm m\,s^{-1}$ between the Keplerian fit of radial velocities secured over more than two years (Figure \ref{fig: xo6 RVs}) and the spectral transit observed during a single night (Figure \ref{fig: xo6 rossiter}). This shift is of the order of magnitude of the Keplerian fit residuals dispersion so it is not significant. This shows that the systematics and stellar jitter are smaller on a few-hour timescale than on a two-year time scale.}

\begin{table}
\begin{center}
\caption{Radial velocities of XO-6 measured along the orbit and during the transit of January 17, 2015 using the SOPHIE spectrograph at the Observatoire de Haute-Provence.}
\label{tab: xo6 RVs}
\begin{tabular}{cccc}
\hline\hline
Reduced  & Orbital   & RV  & 1-$\sigma$ uncertainty  \\
BJD & phase &  [$\rm km\,s^{-1}$] &  [$\rm km\,s^{-1}$]  \\
\tableline 

56551.6553  &   0.1588   &  -5.965  &  0.109  \\ 
56560.6592  &  -0.4497   &  -5.394  &  0.114  \\ 
56560.6617  &  -0.4491   &  -5.574  &  0.100  \\ 
56723.3858  &  -0.2289   &  -5.090  &  0.089  \\ 
56974.5451  &   0.4801   &  -5.642  &  0.090  \\ 
57040.3717  &  -0.0361   &  -5.236  &  0.089  \\ 
57040.3814  &  -0.0335   &  -5.143  &  0.085  \\ 
57040.3916  &  -0.0308   &  -5.265  &  0.080  \\ 
57040.4017  &  -0.0281   &  -5.153  &  0.083  \\ 
57040.4114  &  -0.0255   &  -5.252  &  0.081  \\ 
57040.4209  &  -0.0230   &  -5.109  &  0.086  \\ 
57040.4300  &  -0.0206   &  -5.196  &  0.089  \\ 
57040.4390  &  -0.0182   &  -5.279  &  0.085  \\ 
57040.4486  &  -0.0156   &  -5.182  &  0.087  \\ 
57040.4585  &  -0.0130   &  -5.034  &  0.087  \\ 
57040.4683  &  -0.0104   &  -4.958  &  0.087  \\ 
57040.4784  &  -0.0077   &  -5.171  &  0.089  \\ 
57040.4885  &  -0.0051   &  -5.241  &  0.095  \\ 
57040.4986  &  -0.0024   &  -5.521  &  0.087  \\ 
57040.5085  &   0.0003   &  -5.473  &  0.086  \\ 
57040.5181  &   0.0028   &  -5.652  &  0.086  \\ 
57040.5272  &   0.0052   &  -5.706  &  0.083  \\ 
57040.5364  &   0.0077   &  -5.664  &  0.090  \\ 
57040.5460  &   0.0102   &  -5.533  &  0.089  \\ 
57040.5561  &   0.0129   &  -5.691  &  0.089  \\ 
57040.5657  &   0.0155   &  -5.533  &  0.079  \\ 
57040.5746  &   0.0178   &  -5.239  &  0.088  \\ 
57040.5840  &   0.0203   &  -5.420  &  0.096  \\ 
57363.6254  &  -0.1785   &  -5.428  &  0.087  \\ 
57378.4922  &  -0.2298   &  -5.249  &  0.080  \\ 
57383.4730  &   0.0931   &  -5.554  &  0.094  \\ 
57384.5389  &   0.3762   &  -5.315  &  0.083  \\ 
57399.3742  &   0.3165   &  -5.603  &  0.112  \\ 
57402.3977  &   0.1196   &  -5.663  &  0.083  \\ 
57402.6758  &   0.1934   &  -5.576  &  0.099  \\ 
57405.5191  &  -0.0514   &  -5.426  &  0.109  \\ 

\hline\hline
\end{tabular}
\end{center}
\vspace{-2mm}
Note. The orbital phase is 0 at mid-transit.
\end{table}

\subsection{Doppler tomography analysis}
\label{sec: Doppler tomography analysis}

\begin{figure}[tb]
\includegraphics[angle=0,width=\columnwidth]
{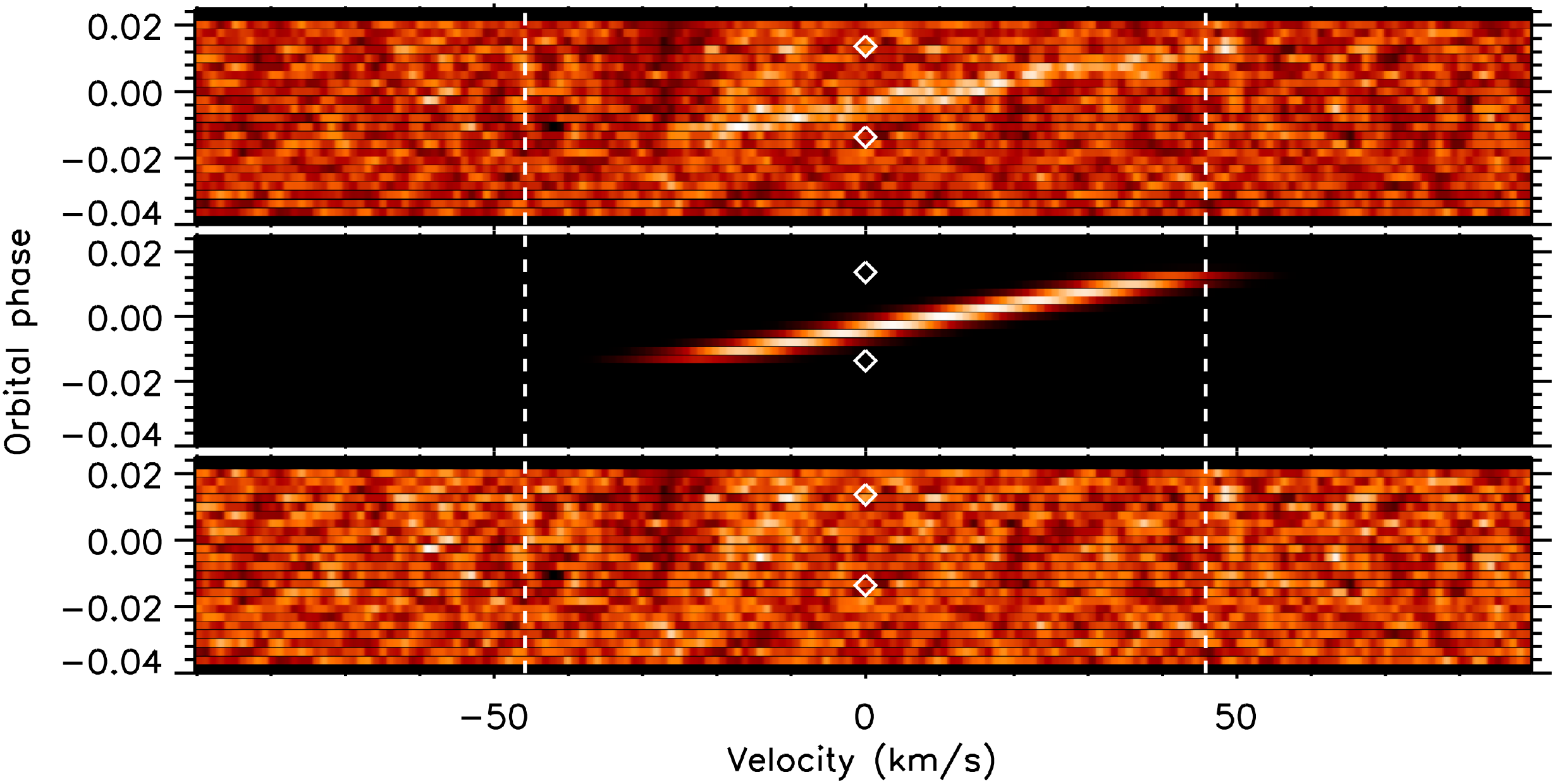}
\caption[]{
Maps of the time-series CCFs during the transit of January 17, 2015, as a function of the radial velocity relative to the star 
(in abscissa) and the orbital phase (in ordinate, increasing vertically). Flux values increase from dark red to white. 
Vertical dashed white lines are plotted at $\pm$ $v$sin$i_{\star}$, and the white diamonds indicate the time of the first 
and the fourth contacts. 
\textit{Top panel}: Map of the transit residuals after subtracting the \modif{modeled out-of-transit CCF}. The signature of XO-6b is a bright and wide feature 
that crosses the line profile from $\sim$$-20$\,km\,s$^{-1}$ at the ingress to $\sim$$+40$\,km\,s$^{-1}$ at the egress. \modif{The dark feature at -30$\rm \,km\,s^{-1}$ probably arises from a stellar pulsation}.
\textit{Middle panel}: Best-fit model for the XO-6b transiting signature. The parameters of this best-fit model are 
constrained by both the spectroscopic and the photometric data. 
\textit{Bottom panel}: Overall residual map after the subtraction of the model planet signature. \vspace{1mm}
\label{Fig:tomography}}
\end{figure}

We analyzed the spectra obtained during the transit 
using a line-profile tomography technique as described in \citet{Collier2010a}. This technique is particularly well suited to fast 
rotators \citep{Collier2010b, Bourrier2015}. 
The planet partial occultation of the star produces a Gaussian ``bump'' in the stellar line profile, 
which spectral location depends on the planet position in front of the stellar disk 
during the transit (Figure~\ref{Fig:tomography}).
We fit the CCFs of the stellar spectra with a transit model, with and without prior constraints 
obtained previously from the photometric transits. 
%

%
%
We use the spectra before and after the transit to measure 
the fixed pattern in the CCF \citep[the so-called ``side-lobes pattern'',][]{Collier2010a}, 
which is assumed to be constant over the same night and are 
only function of the velocity point $j$ in the CCF abscissa. 
The fixed pattern is first estimated by the mean of the residual in the out-of-transit spectra 
after a fit to the data by the model of the stellar line. We then simultaneously 
search for the fixed pattern and the best star+transit model in a global fit.
%
%
The free parameters of the fit are the systemic velocity $v_{\star}$, the star projected rotational velocity 
$v\sin i_{\star}$, the system size $a/R_{\star}$, the planet-to-star size ratio $R_p/R_{\star}$, 
the inclination of the planetary orbit $i$, the sky-projected obliquity $\lambda$, and 
the local line profile width $s$=$\sigma/(v\sin i_{\star})$ 
expressed in units of the projected stellar rotational velocity, 
where $\sigma=FWHM/(2\,\sqrt{2\,ln\,2})$ and $FWHM$ 
is the full-width at half-maximum of the non-rotating stellar line profile convolved 
with the instrument profile \citep[see details in][]{Bourrier2015}.
The orbital period and the central time of the transit are taken from the analysis of the photometric data. 
%
%
We performed a fit with and without the constraints on the two free parameters measured 
independently from the photometric data ($a/R_{\star}$ and $i$).
%
%
%
%
The merit function of the fit is a $\chi^2$ function as in \citet{Bourrier2015}:
\begin{equation}
\begin{split}
\chi^2=&\sum\limits_{i}^{n_{CCF}} \sum\limits_{j}^{n_v} \left[ \frac{f_{i,j}(model)-f_{i,j}(obs)}{\sigma_{i}}  \right]^2     + \\
&\sum\limits_{a_\mathrm{p}/R_{\star}, i_\mathrm{p}}\left[ \frac{x_{tomo}-x_{photo}}{\sigma_{x_{photo}}}  \right]^2  +
\left( \frac{v_{\star}- v_{\star,obs}}{\sigma_{v_{\star,obs}}} \right)^2,
\end{split}	
\label{eq:stdev}  
\end{equation}
where $f_{i,j}$ is the flux at velocity point $j$ in the $i$th observed or model CCFs. 
The error on the flux, $\sigma_{i}$, is supposed constant for a given CCF. 
The photometric measurements of the semi-major axis and orbital inclination ($x_{photo}\pm\sigma_{x_{photo}}$) can be used to constrain the model values of these parameters $x_{tomo}$ (Table~\ref{Tab:Tomography}). 
The observed star systemic velocity $v_{\star,obs}$ is taken to be -5.1\,km/s 
with a conservative error bar of $\sigma_{v_{\star,obs}}$=1.0\,km/s. \modif{The difference with the systemic velocities 
reported in Section \ref{sec: Radial velocity follow-up} is due to the distinct methods used to determine them
from the CCF fits (a simple Gaussian fit in Section \ref{sec: Radial velocity follow-up} and a more elaborate 
profile fit here). These differences have no effects on the results.}
%
%

%
%
For the tabulated errors on the CCF profiles $\sigma_i$,
we use the residuals between the CCFs and the first best-fit model profile 
obtained assuming the same error bars for all CCF. 
We also took into account that the CCFs are calculated by the instrument pipeline
at a velocity resolution of 0.5\,km\,s$^{-1}$, while the spectra have an instrumental resolution 
of about 7.5\,km\,s$^{-1}$. 
The residuals are thus strongly correlated, which can lead to an underestimation of the error bars 
on the derived parameters. As in \citet{Bourrier2015} we retrieved the uncorrelated Gaussian component 
of the CCFs noise (the ``white noise'') using an analysis of the residuals variance as a function of a 
data binning size $n_{bin}$. 
The variance well matches the quartic harmonic combination of a white and a red noise components:
\begin{equation}
\sigma^2(n_{bin})=\left(\left(\frac{n_{bin}}{\sigma_{Uncorr}^2}\right)^{2}+\left(\frac{1}{\sigma_{Corr}^2}\right)^{2}    \right)^{-\frac{1}{2}}.  
\label{eq:stdev}   			
\end{equation}
where $\sigma_{Uncorr}/\sqrt{n_{bin}}$ can be understood as the result of the intrinsic uncorrelated noise 
after the binning of $n_{bin}$ pixels, 
and $\sigma_{Corr}$ is a constant term characterizing the correlation between the binned pixels. 
%
%
We find that the uncorrelated noise is about 
$\sqrt{2}$ larger than the dispersion of the non-binned data. 
For the final fit, we did not bin the data, 
and for all CCF pixels we use the uncorrelated noise for the tabulated errors $\sigma_i$. 

The result of the fit to the spectroscopic profiles of the transit  
is summarized in \modif{Table~\ref{Tab:Tomography}.} 
%
%
%
%
The $v\sin i_{\star}$ is measured to be 48$\pm$3\,km\,s$^{-1}$.
%
%
%
%
The transit of the planet is clearly detected in the CCF profiles (Figure~\ref{Fig:tomography}).
The effect of the dark silhouette of the planet is a bump superimposed on the stellar absorption line, 
which goes from about $-20$\,km\,s$^{-1}$ at the ingress, to about $+40$\,km\,s$^{-1}$ 
at the egress, both relative to the systemic velocity of the star. This asymmetry is a signature of a prograde orbit 
that is misaligned to the equatorial plane of the star: 
the ingress of the planet shadow is at larger latitude on the stellar disk than the egress. 
The spectroscopic detection of the planet shadow yields an estimate of the planet size ($R_p/R_{\star}=0.107^{+0.006}_{-0.005}$), 
which is totally independent from and consistent with the estimate from time-series photometry. This shows that the transiting object detected by the tomographic analysis and from the photometric lightcurves is indeed the same, and indeed transits in front of \xos.
%
%
The combined measurements using both photometric estimate and spectroscopy data gives $R_p/R_{\star}$=0.110$\pm$0.006.
%
%
%
%

Finally, the planet orbit misalignment is robustly established by the asymmetry of the 
Rossiter-MacLaughin radial velocity anomaly seen in the radial velocity curve during the transit (Figure~\ref{fig: xo6 rossiter}) and the Doppler position of the planet shadow on the stellar disk seen in the CCF profiles (Figure~\ref{Fig:tomography}). 
%
%
%
%
We obtain a measurement of the sky-projected obliquity that is at 9-$\sigma$ different from zero: 
$\lambda$=$-$20.7$\pm$2.3\,degrees. 
%
%
%
%

\begin{table*}
\begin{center}
\caption{
Results of the Doppler tomography, with and without the constraints 
from the photometric data. 
The 3$^{\rm rd}$ column reports the values obtained with the analysis of the photometric data only.
The 4$^{\rm th}$ column, labeled Tomography, gives the results obtained using the spectroscopic data only. 
%
%
%
%
The 5$^{\rm th}$  column gives the results obtained using both 
the fit to the spectroscopic data and the additional constraints 
of the two first values (scaled semi-major axis and orbital inclination) 
from the photometry as tabulated in the 3$^{\rm rd}$  column.
%
%
%
%
\label{Tab:Tomography}
}
\begin{tabular}{lccccl}
\hline
\noalign{\smallskip}  
{Parameter}   & {Symbol} 	& {Photometry}       & {Tomography}              & {Tomography}		   & {Unit} \\   			
                     &                 	&                           &                                  & {+ Photometry}		   &  \\   			
\noalign{\smallskip}
\hline
\hline
\noalign{\smallskip}
Scaled semi-major axis 		& $a/R_{\star}$	& 9.20$\pm$0.19  		  &8.3$\stackrel{+1.2}{_{-0.8}}$						&9.08$\pm$0.17	& -- \\
Orbital inclination  	 & $i$ 		          & 85.95$\pm$0.20 			&85.4$\stackrel{+1.5}{_{-1.0}}$						&86.0$\pm$0.2	& deg \\
Planet-to-star radii ratio&$R_\mathrm{p}/R_{\star}$   & 0.1144$\pm$0.0013		&0.107$\stackrel{+0.006}{_{-0.005}}$							            &0.110$\pm$0.006								&  -- \\
Stellar rotation velocity & $v$sin$i_{\star}$         & -										&51$\pm$5 		   						              &48$\pm$3        								&	km\,s$^{-1}$ \\
Local line width ratio 	& $s^{\dagger}$									& -								
		&0.120$\stackrel{+0.016}{_{-0.013}}$  			&0.129$\pm$0.013&  -- 	\\
%
%
%
%
Local line width 
%
%
%
 			& $s\times v\sin i_{\star}$									& -								
		&6.1$\pm$ 0.3  			&6.2$\stackrel{+0.5}{_{-0.4}}$&  km\,s$^{-1}$ 	\\
Sky-projected obliquity 	& $\lambda$ 								& -										&$-18.1$$\stackrel{+3.4}{_{-8.0}}$  		 					&$-20.7$$\pm 2.3$ 	  &	deg\\      
\hline									
\hline
\end{tabular} \\
\end{center}
\vspace{-3mm}
\hspace{1.35cm} $\dagger$: $s$=FWHM/(2\,$\sqrt{2\,ln\,2}$\,$v$sin$i_{\star}$).
\end{table*}

\subsection{The planetary nature of \xosb}

Because of the fast stellar rotation, the recoil radial velocity amplitude and mass determination of \xosb are uncertain. In the following, we review the possible configurations for the \xos system and assess the planetary nature of the transiting object through the elimination process, as done for KOI-12b (Kepler-448b) \citep{Bourrier2015}.

\begin{itemize}
\item The radial velocity measurements taken during the orbit provide a 3-$\sigma$ upper limit for the mass of the companion of $M_p < 4.4 \: \rm M_{Jup}$. Therefore, we can exclude the case of an unblended stellar eclipsing binary system. 

\item An eclipsing binary in the background of \xos could mimic the photometric transit signatures. However, the detection of the Rossiter-McLaughlin effect and the tomographic analysis indicate that the transit occurs in front of the fast rotating star, which rules out a background eclipsing binary as the source of the transits.

\item If \xos is an eclipsing binary system diluted in the light of third star, this star might contaminate the transit signatures and stellar line profile distortions. However, no indication of another star is seen in the spectra. If such a star is present but too faint to be detected, then it would not affect significantly the analysis nor the planetary nature of \xosb. In addition, the transit parameters extracted from the photometry and from the tomographic analysis are consistent with each other. This argues against the case of a third star that would contaminate the photometric transit signature without affecting the spectroscopic measurements, and vice-versa. If a third star is present and bright enough to affect all measurements consistently, it should have the same spectral type, same systemic radial velocity, same rotation rate, and same spin alignment as \xos to remain undetected in the spectra and in the tomographic analysis; this contrived identical twin stars configuration is extremely improbable.

\end{itemize}

Overall, the ensemble of data allows us to validate the planetary nature of \xosb. Its parameters are given in Table \ref{tab: final parameters}.

\begin{table}[htdp]
\begin{center}
\caption{Parameters of the XO-6 system assuming a circular orbit.}
\label{tab: final parameters}
\begin{tabular}{cccc}
\hline
\hline
   Parameter &  Units & Value & 1-$\sigma$ uncertainty   \\
\hline

$P$     & [days]   &  3.7650007  &  0.0000081  \\
$T_0$ & [BJD]    &  2456652.71245  &  0.00055  \\
$a/R_{\star}$    &    &  9.08  &  0.17  \\
$R_p/R_{\star}$    &    &      0.110     &   0.006  \\
$i$    & [deg]   &      86.0    &   0.2 \\
$\tau_{14}$  & [hours]   &  2.90    &  0.10   \\
$\tau_{23}$  & [hours]   &  1.99    &  0.12    \\

$b$   &    &    0.633    &  0.034   \\
$a$   & [au]     &	      0.0815	&   0.0077	  \\ 
$\rho_{\star a}$$^{\dagger}$  & [g cm$^{-3}$]   &  1.0  &  0.056 \\ 
$K_{\star}$ & [m s$^{-1}$] &  200  & 70 \\     
		 	&		 	&  $< 450$  & \\     
$M_p$     & [$\rm M_{Jup}$]   & 1.9  & 0.5  \\      
		    &				   & $< 4.4$  &  \\      
$R_p$$^{\dagger}$     & [$\rm R_{Jup}$]   & 2.07  &  0.22  \\      
$\rho_p$  & [g cm$^{-3}$]   &  0.27  &  0.11 \\ 
		& 			   &  $<$ 0.62 &  \\ 
$g_p$   & [m s$^{-2}$]     &   26.4  &  9.7   \\
		&			     &   $< 59.4$   &     \\
$H$  &  [km]  &  216  &  80  \\      
	 & 		  &  $> 96$  &  \\      
$T_{eq}$  &  [K]  & 1577 &  28 \\       

\hline  
\hline  
\end{tabular}
\end{center}
\vspace{-2mm}
Notes. $^{\dagger}\,$See Section \ref{sec: Properties of the hot Jupiter xosb} for a discussion on the planetary radius and the stellar density.
$<$ indicate 3-$\sigma$ upper limits, $>$ indicate 3-$\sigma$ lower limits.
\end{table}

\section{Stellar parameters}
\label{sec: Stellar parameters}

\begin{figure}[htbp]
   \centering
   \includegraphics[width=6.5cm]{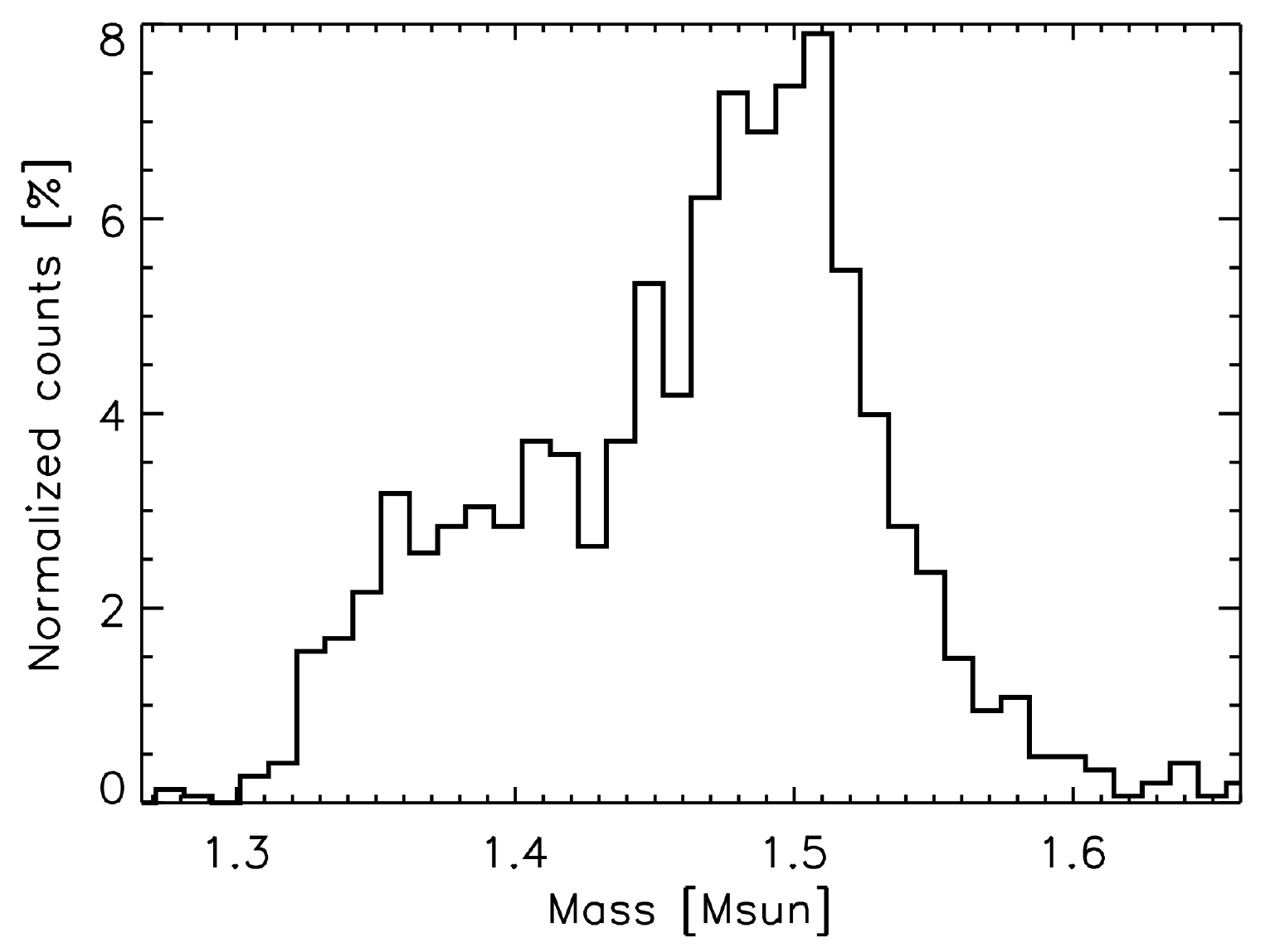}
   \includegraphics[width=6.5cm]{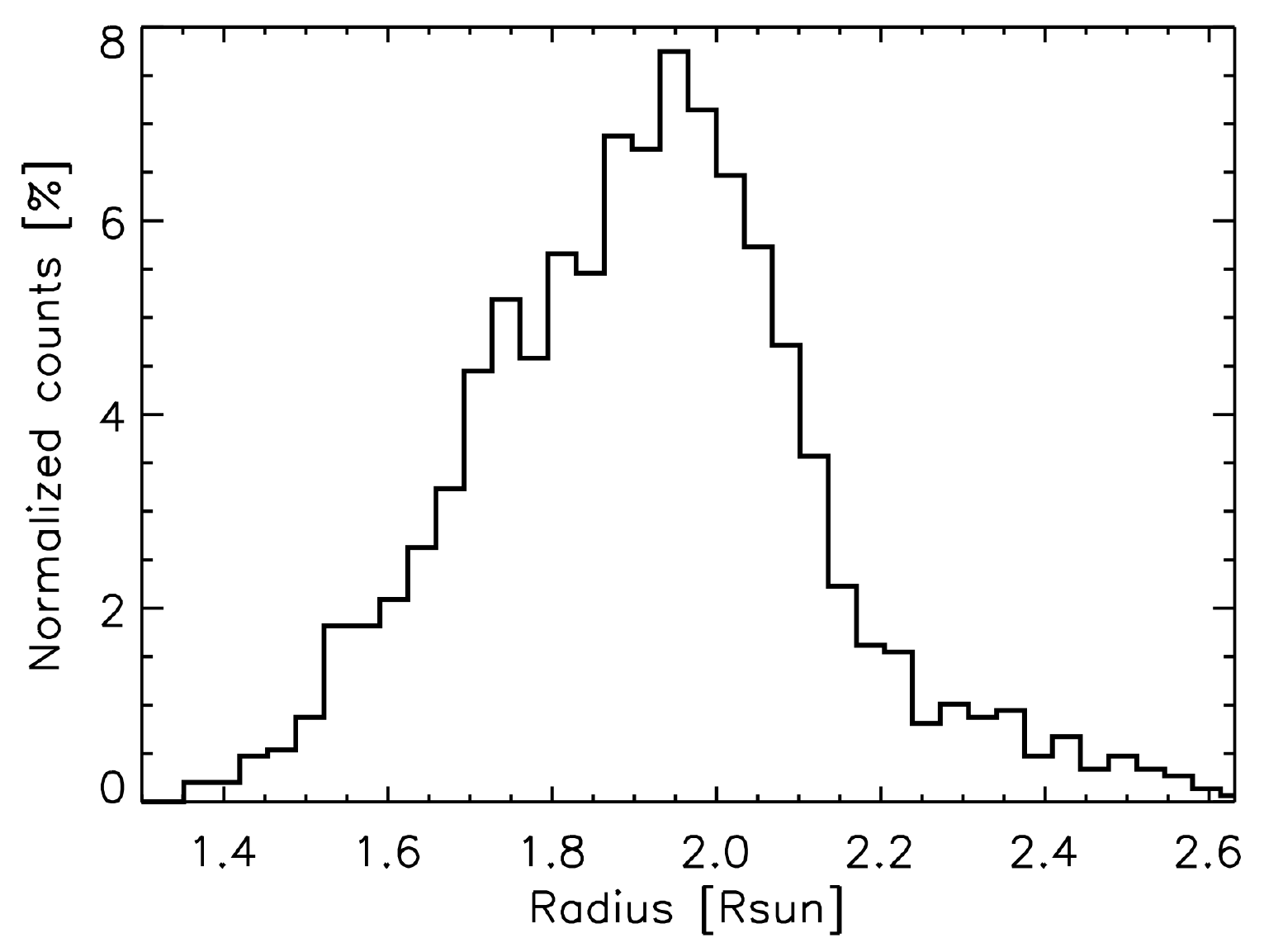}
   \includegraphics[width=6.5cm]{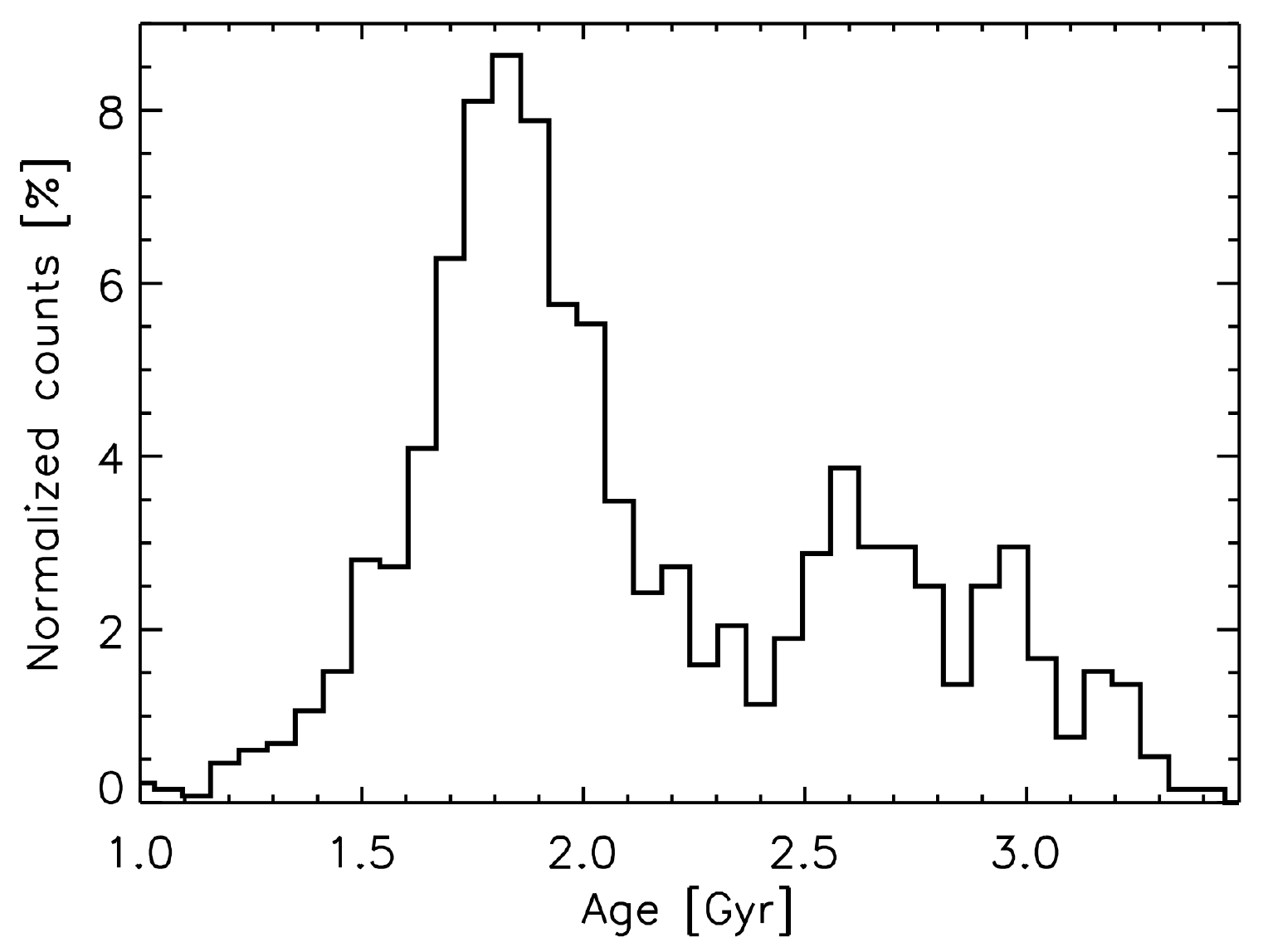}
    \caption{Distributions of the mass, radius, and age for the \xos star.}
   \label{fig: stellar param}
\end{figure}

We derive the stellar properties from a spectral analysis of SOPHIE spectra. We keep only spectra with a signal to noise ratio larger than 40, and co-add them once corrected from the radial velocities and set in the rest frame. The final spectrum has a signal to noise ratio of \modif{\simi640 per resolution element in the continuum at 5610 $\rm \AA$}. However, the fast stellar rotation yields wide and blended spectral lines which complicate the analysis. First, we derive $v\,$sin$\,i_{\star}$ and the macroscopic velocity $v_{macro}$ by analyzing isolated spectral lines in the Fourier space. Then, we use the SME package \citep{Valenti1996} to fit the observed spectrum to a synthetic spectrum. We use reference abundances from \citet{Asplund2005}. We derive the effective temperature $T_{eff\star}$ on the hydrogen lines, the gravity log$g_\star$ on the Ca I and Mg I lines, and the metallicity $[Fe/H]$, all of which through an iterative process. The results are reported in Table \ref{tab: stellar properties}. The values for $T_{eff\star}$ and $[Fe/H]$ are compatible within their uncertainties with those reported by \citet{Ammons2006}. 
Using the atmospheric parameters $T_{eff\star}$, log$g_\star$, and $[Fe/H]$ from the spectral analysis, we estimate the mass, radius, and age by a MCMC minimisation procedure using the STAREVOL stellar evolution models \citep{Siess2000, Palacios2003, Palacios2006, Decressin2009, Lagarde2012}. The atmospheric parameters are consistent with main sequence evolutionary tracks, and marginally consistent with pre-main sequence tracks. We exclude the pre-main sequence cases because the star does not show any sign of a young age. We also exclude the post-main sequence cases as they are outside of the 1-$\sigma$ uncertainties of $T_{eff\star}$ and log$g_\star$. Finally, our models do not include overshooting, but for \xos the major source of uncertainty arises from the stellar rotation.
The posterior distributions for the stellar age, mass, and radius are shown in Figure \ref{fig: stellar param} and the corresponding parameters are reported in Table \ref{tab: stellar properties}. The distributions are bimodal for the age and mass ; we also report in Table \ref{tab: stellar properties} the solutions that correspond to these two modes taken individually (solutions a and b).
Using the stellar density obtained from the photometric $a/R_{\star}$ instead of the spectroscopic log$g_{\star}$ would yield different solutions for the stellar age, mass, and radius. However, because the orbital solution is poorly constrained, we keep log$g_\star$ as an input for the stellar evolution models (see Section \ref{sec: Properties of the hot Jupiter xosb} for a discussion on the stellar density).

\begin{table}[htdp]
\begin{center}
\caption{Properties of the XO-6 star.}
\label{tab: stellar properties}
\begin{tabular}{cccc}
\hline
\hline
   Quantity &  Unit &  Value  & Notes   \\
\hline

Name  &  &  TYC 4357-995-1   &  1   \\
RA   & J2000 & 06:19:10.3604 &  1   \\
DEC  & J2000 &  +73:49:39.602  &  1    \\
V & Mag  &  10.25 $\pm$ 0.03 &  2   \\
H  & Mag  & 9.266 $\pm$ 0.017 &  2   \\
B-V & Mag  & 0.43 $\pm$ 0.04 &  2   \\
Distance & pc & $86^{+79}_{-31}$  &  3  \\
               &      & $196$                    &  4   \\
Spectral type &  & F5 &  2, 5   \\
$T_{eff\star}$ & K  &  $6720.00 \pm 100$ &  5 \\
$[Fe/H]$ &   & $-0.07 \pm 0.1$             &  5    \\
log$g_{\star}$ 		&	& $4.04 \pm 0.10$		& 5   \\
$v_{micro}$	&  $\rm km\,s^{-1}$  & $1.95 \pm 0.5$  & 5 \\
$v_{macro}$	&  $\rm km\,s^{-1}$  & $20.6 \pm 4.5$  & 5 \\
$v\,$sin$\,i_{\star}$ 	&  $\rm km\,s^{-1}$  & $44.2 \pm 3.0$  & 5 \\
Mass	& M$_{\odot}$	& $1.47 \pm 0.06$	& 5  \\
Radius	& R$_{\odot}$	& $1.93 \pm 0.18$	& 5  \\
Age		& Gyr			& $1.88^{+0.90}_{-0.20} $	& 5  \\
\hline
Mass	& M$_{\odot}$	& $1.50 \pm 0.03$	& 5, a \\
Radius	& R$_{\odot}$	& $1.86 \pm 0.16$	& 5, a \\
Age		& Gyr			& $1.86 \pm 0.20$	& 5, a \\
Mass	& M$_{\odot}$	& $1.39 \pm 0.04$	& 5, b  \\
Radius	& R$_{\odot}$	& $2.05 \pm 0.12$	& 5, b  \\
Age		& Gyr			& $2.74 \pm 0.32$	& 5, b  \\
\hline
\hline  
\end{tabular}
\end{center}
Notes. 1: \citet{Hog1998}. 2: \citet{Hog2000}. 3: \citet{Ammons2006}. 4: \citet{Pickles2010}. 5: This work. a, b: Solutions for each of the two modes (see text and Figure \ref{fig: stellar param}).
\end{table}

\section{Discussion}
\label{sec: Discussion}

\subsection{Properties of the hot Jupiter \xosb}
\label{sec: Properties of the hot Jupiter xosb}

The orbits of hot Jupiters are expected to be circularized \citep{Leconte2010}. However, several hot Jupiters have been found on eccentric orbits such as XO-3b ($e = 0.26$), HAT-P-34b ($e = 0.44$), HAT-P-2b ($e = 0.52$), or CoRoT-20b ($e=0.56$). In the case of \xos, the stellar rotation may excite the planetary orbit and prevent its circularization, as discussed in Section \ref{sec: Stellar rotation and planetary orbits}. However, without a constraint on the eccentricity due to the difficulty of measuring radial velocities on this fast rotating star, we derive the planet parameters assuming a circular orbit.
Using the stellar radius derived in Section \ref{sec: Stellar parameters}, we find a planetary radius $R_p=2.07 \rm \; R_{Jup}$. Assuming a mass $M_p=1.9 \rm \; M_{Jup}$ suggested by the radial velocities and a resulting density $\rho_p=0.27 \; \rm g\,cm^{-3}$, we find that \xosb is extremely inflated compared to model predictions for hot Jupiters. Based on models by \citet[][see also \citealt{Moutou2013}]{Guillot2006}, \xosb appears to have a positive radius anomaly of about $+65000$~km that is the largest of known exoplanets. For comparison, another extremely inflated planet, \mbox{CoRoT-2b}, has a radius anomaly of about $+30000$~km which already cannot be explained by standard recipes \citep[\eg][]{Guillot2011}. However, this extremely large $R_p$ value might be due to a poor estimate of $R_{\star}$. Using $\rho_{\star a}$ instead of log$g_{\star}$ as an input for the stellar evolution models, the stellar radius would decrease to $1.21 \rm \; R_{\odot}$ and would yield a planetary radius of $1.30 \rm \; R_{Jup}$. Although still inflated, this would be in line with other inflated hot Jupiters.

Using Equation 4 of \citet{Southworth2007} with $e = 0$ and $K_{\star} = 200 \rm \; m\,s^{-1}$, we derive a planet's surface gravity $g_p = 26 \rm \; m\,s^{-2}$ . Assuming a zero albedo, we derive an equilibrium temperature $T_{eq} = 1577$ K and a scale height $H = 216$ km. This temperature is close to the condensation temperature of species thought to be responsible for clouds in the atmospheres of hot Jupiters, such as $\rm MgSiO_3$ \citep{Lodders1999, Lecavelier2008, Pont2013}. \xosb is warm compared to other close-in gas giant planets for which molecular spectral signatures have been detected, such as HD~209458b, XO-1b, and HD~189733b \citep{Deming2013, McCullough2014, Crouzet2014}, and it orbits a bright star (H = 9.27). \modif{In addition, the nearby star located at only 38'' separation and only 1 magnitude fainter may provide an excellent reference.} Thus, \xosb is well suited to atmospheric studies, and the \xos system is untypical because of the high temperature of the host star.

Because the star is a fast rotator, measuring the mass of \xosb is challenging. A few other transiting hot Jupiters have been found around such hot and fast rotating stars (Table~\ref{tab: fast rotating systems}). CoRoT-11, WASP-33, and KELT-7 have been observed extensively in radial velocities, with 31, 248, and 36 measurements along the orbit respectively, yielding the planets' masses \citep{Gandolfi2010, Lehmann2015, Bieryla2015}. CoRoT-3b is so massive that the recoil motion overwhelms stellar rotation effects \citep{Deleuil2008}. Only a mass upper limit has been inferred for KOI-12 \citep{Bourrier2015}. Kepler-13Ab has its mass determined from the Kepler photometric lightcurves \citep[][see also \citet{Santerne2012}]{Shporer2011, Mazeh2012, Mislis2012, Shporer2014}. \modif{HAT-P-56 is slightly less massive than \xos and more radial velocity data were obtained, and the KELT-17 radial velocity data show less scatter than for \xos \citep{Huang2015, Zhou2016}}. For \xos, we collected 14 spectra along the orbit and we detect a hint of variation, yielding a mass of 1.9$\,\pm\,0.5\, \rm M_{Jup}$ and a 3-$\sigma$ upper limit of 4$\,\rm M_{Jup}$ (Figure \ref{fig: xo6 RVs}); an extensive radial velocity campaign would be necessary to measure the planet's mass with greater confidence. 
An alternative approach would be to derive the mass from spectroscopy using a method suggested by \citet{deWit2013}, in which the planet's mass is obtained from the atmospheric Rayleigh scattering slope and the atmospheric temperature:
\begin{equation}
M_p = - \frac{4kT[R_p(\lambda)]^2}{\mu G \frac{\text{d}R_p(\lambda)}{\text{d\,ln}\,\lambda}}
\label{eq: mass}
\end{equation}
where $k$ is the Boltzmann constant, $T$ the planet's atmospheric temperature, $\mu$ the atmospheric mean molecular mass, $G$ the gravitational constant, and $\lambda$ the wavelength. The Rayleigh scattering slope $\frac{\mathrm{d} R_p(\lambda)}{\mathrm{d\,ln}\,\lambda}$ could be obtained by transmission spectroscopy with \textit{HST} STIS and ACS, and the atmospheric temperature could be inferred by measuring the eclipse depth with \textit{Spitzer}. Although expensive in terms of resources, these measurements would also characterize the atmosphere of \xosb.

\begin{table}[htp]
\begin{center}
\caption{Hot Jupiter systems with $v\,$sin$\,i_{\star} > 30 \rm \; km \, s^{-1}$ or $T_{eff\star} > 6700 \rm \; K$, and a measured sky-projected obliquity.}
\label{tab: fast rotating systems}
\begin{tabular}{ccccc}
\hline
\hline
Name  & $T_{eff\star}$ &  $v\,$sin$\,i_{\star}$   & $\lambda$ 	& $M_p$  				   \\
	     & [K]	    	    & [$\rm km \, s^{-1}$]	 & {[deg]} 	&	{[M$\rm_{Jup}$]}		\\	     
\hline
CoRoT-11		&	6440$\pm120$	&	40$\pm5$	&	0.1$\pm2.6$	&	2.33$\pm0.27^a$			\\
{HAT-P-56}		&	6566$\pm50$	&	36.4$\pm0.7$	&	8$\pm2$	&	2.18$\pm0.25^b$	\\
\xos				&	6720$\pm100$	&	48$\pm3$	&	
{-20.7$\pm$2.3}
		&	$< 4.4^c$ 					\\
CoRoT-3		&	6740$\pm140$	&	17$\pm1$	&	-37.6$^{+22.3}_{-10}$	&	21.77$\pm1.0^d$			\\
KELT-7 			&	6789$^{+50}_{-49}$	&	69.3$\pm{0.2}$	&	2.7$\pm0.6$		&	1.28$\pm0.18^e$			\\
KOI-12			&	6820$\pm120$	&	60$^{+0.9}_{-0.8}$	&	12.5$^{+3}_{-2.9}$	&	$< 10^f$				\\
WASP-33		&	7430$\pm100$	&	90$\pm10$	&	-108.8$\pm1$	&	2.1$\pm0.2^g$ 				\\
{KELT-17}		& 	7454$\pm49$ &	44.2$^{+1.5}_{-1.3}$		&	-116$\pm4$	&	1.31$\pm0.29^h$	\\
Kepler-13A		&	7650$\pm250$	&	76.96$\pm0.61$	&	58.6$\pm2$		&	6.52$\pm1.57^i$				\\

\hline  
\hline  
\end{tabular}
\end{center}
Reference for $M_p$: a: \citet{Moutou2013}. b: \citet{Huang2015}. c: This work. d: \citet{Deleuil2008, Moutou2013}. e: \citet{Bieryla2015}. f: \citet{Bourrier2015}. g: \citet{Lehmann2015}. h: \citet{Zhou2016}. i: \citet{Shporer2011}.
\end{table}

The radial velocities do not constrain the eccentricity and we assumed $e$ = 0 in our analysis. Under this assumption, the stellar density obtained from the photometric transit curve is $\rho_{\star a} = \frac{3\pi}{GP^2}\,(\frac{a}{R_{\star}})^3 = 1 \pm 0.056 \, \rm g\,cm^{-3}$. This density is 1-$\sigma$ away from the closest evolution track for these $T_{eff\star}$ and $Fe/H$. In addition, it differs significantly from that obtained from the mass and radius of Table \ref{tab: stellar properties}: $\rho_{\star b} = \frac{M_{\star}}{\frac{4}{3} \pi R_{\star}^3} = 0.29 \pm 0.08 \rm \, g\,cm^{-3}$. This discrepancy may indicate a non-zero eccentricity, which can be estimated by:

\begin{equation}
\left(\frac{\rho_{\star a}}{\rho_{\star b}}\right)^{1/3} = \frac{1+e\,sin\omega}{\sqrt{1-e^2}} 
\label{eq: ecc}
\end{equation}
where $\omega$ is the argument of periastron \citep[][Eq.~29]{Winn2010b}. This yields $e > 0.39$ with a true value that depends on the unknown value of $\omega$. The radial velocities are still consistent with such eccentricities. Other hot Jupiters orbiting around fast rotating stars have been found on eccentric orbits, as discussed in Section \ref{sec: Stellar rotation and planetary orbits}.
Alternatively, the spectroscopic log$g_{\star}$ is generally poorly constrained compared to $a/R_{\star}$ \citep{Sozzetti2007, Torres2008}; thus, the discrepancy in the stellar density might arise from an inaccurate log$g_{\star}$ as suggested for example in the case of XO-3b \citep{JohnsKrull2008, Winn2008}. 
Another possible estimate is to take the closest density to $\rho_{\star a}$ that is allowed by the evolution tracks for these $T_{eff\star}$ and $Fe/H$. This yields $\rho_{\star c} = 0.88 \pm 0.07 \, \rm g\,cm^{-3}$, which would imply $e > 0.04$. Finally, the discrepancy in $\rho_{\star}$ might be due to a poor estimate of $Fe/H$, or to evolution models that are not adapted to this particular star, or to other issues such as stellar spots \citep[\eg][]{Guillot2011}.

\subsection{Obliquity distribution}

\xosb adds to the sample of hot Jupiters with a measured sky-projected obliquity. We define the sky-projected obliquity as the angle between the stellar spin axis and the normal to the planet's orbital plane, as in \citet{Winn2015}. With the rapid growth of such measurements in recent years, obliquities have become a very promising means of distinguishing between different theories for the dynamical history of hot Jupiters: migration through interactions with the protoplanetary disk or disk-free mechanisms involving interactions with a third body or with other planets \citep[\eg][]{Naoz2011}. In fact, these measurements provide one of the very few observational constraints to these theories. Misalignment of the planet's orbital plane with respect to the stellar spin axis is found to be common: 35\% of known close-in gas giant planets have sky-projected obliquities larger than $30^{\circ}$, and 21\% even have a retrograde orbit. A general picture is emerging: planets orbiting stars with relatively cool photospheres ($T < 6100$ K) have low obliquities whereas planets orbiting hotter stars show a wide range of obliquities \citep{Winn2010a, Schlaufman2010, Albrecht2012, Dawson2014, Winn2015}. The boundary of 6100 K also coincides with the rotational discontinuity above which stars rotate significantly faster \citep{Kraft1967}. In addition, the highest-mass planets ($M_p > 3 \: \rm M_{Jup}$) seem associated with lower obliquities \citep{Hebrard2011}. As an example of theoretical study, \citet{Albrecht2012} could explain qualitatively the observed obliquity distribution as a consequence of tidal timescales, although more parameters such as wind mass loss, stellar evolution, and magnetic braking may also be involved \citep{Valsecchi2014}.

\begin{figure}[htbp]
   \centering
   \includegraphics[width=\columnwidth]{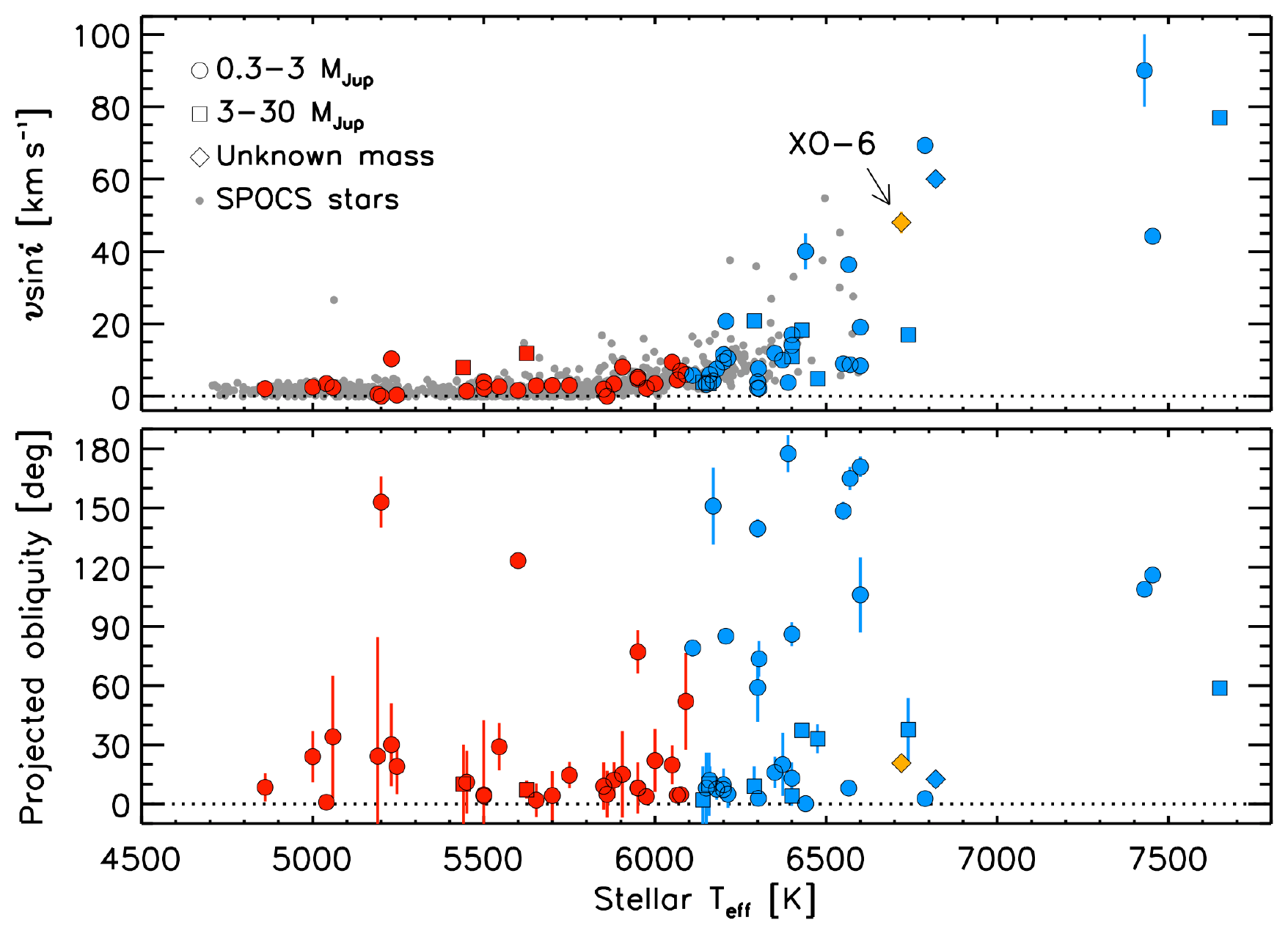}
      \caption{Projected stellar rotation velocity (top) and absolute value of the sky-projected obliquity (bottom) as a function of stellar effective temperature for transiting hot Jupiters ($M_p > 0.3 \:\rm M_{Jup}$, $P < 20$ days) with a measured obliquity. The planet mass is encoded by the symbol shape. Color is used to distinguish temperatures below (red) and above (blue) 6100 K. Grey dots are projected rotation rates of stars in the SPOCS catalog \citep{Valenti2005}. The XO-6 system is shown as a yellow diamond shape. Figure adapted from \citet{Winn2015}, using the databases \url{http://exoplanets.org/} and \url{http://vizier.u-strasbg.fr/}.}
   \label{fig: obliquity distribution}
\end{figure}

Almost all stars hosting a transiting hot Jupiter with a measured sky-projected obliquity have a projected rotational velocity $v\,$sin$\,i_{\star}$ lower than \modif{21}$\rm \; km \, s^{-1}$ (Figure~\ref{fig: obliquity distribution}). \xos lies in the hot star region of the obliquity distribution and bridges the gap between slow and fast rotators. \modif{Five} systems have greater $T_{eff\star}$ and \modif{greater or similar} $v\,$sin$\,i_{\star}$ than \xos: KELT-7, KOI-12, WASP-33, \modif{KELT-17}, and Kepler-13A. CoRoT-3 has a $T_{eff\star}$ similar to \xos but rotates much slower and the transiting object is either a very massive planet or a brown dwarf (Table~\ref{tab: fast rotating systems}).
Among the \modif{nine} systems with $T_{eff\star} > 6700 \rm \; K$ or $v\,$sin$\,i_{\star} > 30 \rm \; km \, s^{-1}$, six of them have low to moderate sky-projected obliquities ranging from 0\degree to 38\degree in absolute value (0\degree to 25\degree if we exclude CoRoT-3b), \modif{one has a large sky-projected obliquity (Kepler-13Ab), and two are in retrograde orbits (WASP-33b and KELT-17b)}. \modif{As a comparison}, for slower rotators between 6100 and 6700~K, the obliquities are widely distributed over the range \mbox{[0\degree, 180$^\circ$]}.

~\\

\subsection{Stellar rotation and planetary orbits}
\label{sec: Stellar rotation and planetary orbits}

\begin{figure}[htbp]
   \centering
   \includegraphics[width=\columnwidth]{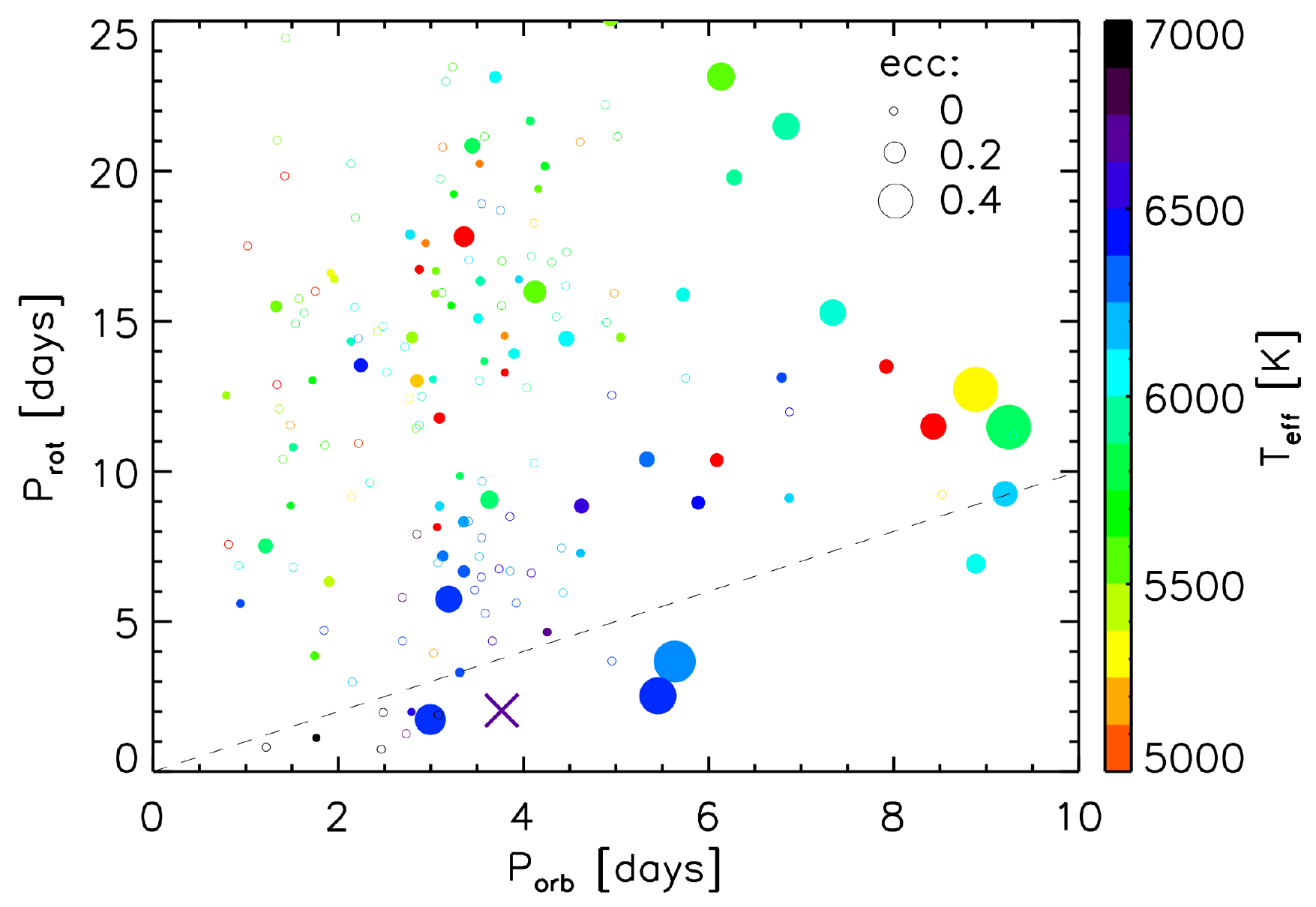}
      \caption{Planetary orbital period $P_{orb}$ and stellar rotation period $P_{rot}$ for hot Jupiters ($M_p > 0.3 \:\rm M_{Jup}$). The $P_{rot}$ are obtained from the stellar $v\,$sin$\,i_{\star}$ and are therefore upper limits, \modif{except in a few cases where they are obtained from photometric modulations}. Colors indicate the stellar effective temperature. Sizes represent the orbital eccentricity; planets with no measured eccentricity are represented by open circles. The dashed line shows the $P_{rot} = P_{orb}$ synchronization. The \xos system is shown as a purple cross, its eccentricity is unknown.}
   \label{fig: prot}
\end{figure}

Studies of relations between stellar rotation and planetary orbital motion also yield constraints on the dynamics of hot Jupiters. Using Kepler data, \citet{McQuillan2013} found a dearth of KOIs (\textit{Kepler} Objects of Interest) at short orbital periods around fast rotating stars: only slow rotators, with rotation periods longer than 5 days, have planets with orbital periods shorter than 3 days. \citep{Teitler2014} attributed this feature to tidal ingestion of close-in planets by their host stars. In addition, several hot Jupiter host stars are in excess rotation, with rotation periods 4 to 8 times smaller than those expected from rotation isochrones, which supports the case for hot Jupiters spinning up their host stars through tidal interactions \citep{Husnoo2012}. A higher magnetic activity has also been found in two hot Jupiter host stars which have excess rotation and for which strong tidal interaction with the planet are expected \citep{Poppenhaeger2014}. 
A general picture of the dynamical evolution of hot Jupiters and their host stars based on tidal interactions is emerging \citep{Husnoo2012,Pont2009,Mazeh2005}. However, this picture applies to the current set of known hot Jupiters, which is almost entirely composed of systems with a stellar rotation period larger than the planet's orbital period ($P_{rot} > P_{orb}$). In fact, this apparent lack of hot Jupiters around fast rotating stars is intriguing. This may be due to an observational bias: these planets are harder to detect or validate by radial velocities because of stellar line broadening, as illustrated in Section \ref{sec: Radial velocity follow-up}, and transit searches tend to target cooler stars. Thus, transiting planet candidates with fast rotating host stars may be missed or rejected more readily. \xosb is one of the very few planets known to orbit a fast rotating star; the stellar rotation period is even smaller than the planet orbital period ($P_{rot} < 2.12$ d, $P_{orb} = 3.77$ d)\footnote{Here, the stellar rotation periods $P_{rot}$ are estimated by $P_{rot} = 2 \pi R_{\star} / v\,$sin$\,i_{\star}$ and are therefore upper limits. The true $P_{rot}$ are smaller by a factor sin$\,i_{\star}$, which is usually unknown.}. Thus, \xos stands in the $P_{rot} < P_{orb}$ domain of the $P_{rot} - P_{orb}$ diagram (Figure \ref{fig: prot}), which corresponds to an interaction regime between hot Jupiters and their host stars that is largely unexplored.

For systems with $P_{rot} < P_{orb}$, tidal forces would be reversed compared to systems with $P_{rot} > P_{orb}$. Tidal forces would raise the planet's angular momentum which would excite its orbit, yielding an increase in orbital period or eccentricity (instead of having the planet spiralling inwards or being circularized on a very close orbit). Over time, tidal dissipation would spin the star down (instead of spinning it up). This tidal push from the star might play a role in the apparent lack of transiting planets at short orbital periods around fast rotators. 
In this context, it is interesting to note that several hot Jupiters lying in the $P_{rot} < P_{orb}$ domain stand out by their large eccentricities: HAT-P-2b ($P_{rot} < 3.66$ d, $P_{orb} = 5.63$ d, $e = 0.52$) and HAT-P-34b ($P_{rot} < 2.52$ d, $P_{orb} = 5.45$ d, $e = 0.44$) which eccentricities have been inferred from radial velocities, and CoRoT-11b ($P_{rot} < 1.73$ d, $P_{orb} = 2.99$ d, $e = 0.35$) which eccentricity has been inferred from the timing offset of the secondary eclipse \citep{Parviainen2013}. The other close-in gas giant planets with $P_{rot} < P_{orb}$ are WASP-33b, \modif{HAT-P-57 b}, OGLE2-TR-L9b, KELT-7b, \modif{HAT-P-56 b, KELT-17 b}, \xosb, and WASP-7b, each of which do not have a measured eccentricity, Kepler-13Ab which may have a very small but non-zero eccentricity \citep{Shporer2014}, CoRoT-6b \modif{which has a moderate eccentricity} ($P_{rot} < 6.92$ d, $P_{orb}= 8.89$ d, $e = 0.18$), and the warm Jupiter KOI-12b (Kepler-448b, $P_{rot} < 1.37$ d, $P_{orb} = 17.86$ d) which does not have a measured eccentricity.
Another interesting case is XO-3b which orbits a relatively fast rotator and shows a relatively large eccentricity ($P_{rot} < 5.75$ d, $P_{orb} = 3.19$ d, $e = 0.29$).
Overall, large eccentricities may be common for planets around fast rotators. Such systems may result in a planet orbit $-$ planet spin synchronization around the pericenter where tidal forces are the strongest, and possibly in stellar spin $-$ planet orbit synchronization also around the pericenter as seems to be the case for CoRoT-11b \citep{Parviainen2013, Moutou2013}.

\section{Conclusion}
\label{sec: Conclusion}

We report a transiting hot Jupiter, \xosb. The fast stellar rotation ($v\,$sin$\,i_{\star} = \modifm{48 \pm 3 \rm \; km\,s^{-1}}$) prevents a characterization and robust mass determination by radial velocities: only a 3-$\sigma$ upper limit is secured on the planet's mass ($M_p < 4.4 \; \rm M_{Jup}$). We confirm the presence of the transiting object using the Rossiter-McLaughlin effect and we analyze the spectra by Doppler tomography. We show that \xosb is a transiting hot Jupiter on a prograde, misaligned orbit with a sky-projected obliquity \modif{$\lambda$=$-$20.7$\pm$2.3}\,degrees. 
\xosb orbits around a bright star (H~=~9.27) and is relatively warm ($T_{eq}=1577$ K), thus it is well suited to atmospheric studies. We derive the stellar parameters through a spectral analysis and a minimization in stellar evolutionary tracks. We find that the host star is a F5 star with $T_{eff\star}= 6720 \pm 100$ K and $[Fe/H]=-0.07 \pm 0.1$, and is consistent with the main sequence. We note a discrepancy between the photometric and spectroscopic values of the stellar density, which may indicate an eccentric orbit or may be due to an inaccurate determination of log$g_{\star}$. This discrepancy also affects the value of the planetary radius.
\xosb adds to the sample of hot Jupiters with a measured sky-projected obliquity, and lies in the hot and fast rotating star region of the obliquity distribution. In addition, the rotation period of the star is smaller than the orbital period of the planet; as a result, tidal forces are reversed compared to hot Jupiters around slow rotators. In this regime, eccentric orbits may be common as suggested by a few striking examples. Overall, the discovery of \xosb provides a new object to study in the context of dynamical interactions between hot Jupiters and their host stars in a parameter space yet largely unexplored.

\acknowledgments
NC acknowledges J. Valenti, H. Neilson, and A. H. M. J. Triaud for useful discussions.
The Dunlap Institute is funded through an endowment established by the David Dunlap family and the University of Toronto.
The XO project is supported by NASA grant NNX10AG30G.
AL, VB, GH, and LA have been supported by an award from the Fondation Simone et Cino Del Duca, and acknowledge the support of the French Agence Nationale de la Recherche (ANR), under program ANR-12-BS05-0012 ``Exo-Atmos''. 
IR, FV and EH acknowledge support from the Spanish MINECO through grant ESP2014- 57495-C2-2-R. The Joan Or\'o Telescope (TJO) of the Montsec Astronomical Observatory (OAdM) is owned by the Generalitat de Catalunya and operated by the Institute for Space Studies of Catalonia (IEEC).
VB work has been carried out in the frame of the National Centre for Competence in Research ``PlanetS'' supported by the Swiss National Science Foundation (SNSF). V.B. acknowledges the financial support of the SNSF.
This research has made use of the Exoplanet Orbit Database and the Exoplanet Data Explorer at exoplanets.org, the Extrasolar Planets Encyclopaedia at exoplanet.eu, and the SIMBAD and VizieR databases at simbad.u-strasbg.fr/simbad/ and http://vizier.u-strasbg.fr/viz-bin/VizieR.
Software: astrometry.net \citep{Lang2010}, Stellar Photometry Software \citep{Janes1993}, JKTLD \citep{Southworth2015}, SME \citep{Valenti1996}, STAREVOL \citep{Siess2000, Palacios2003, Decressin2009, Lagarde2012}, ATLAS \citep{Kurucz1979}.


\end{document}